\documentclass{article}

\usepackage{authblk}
\usepackage[style=numeric-comp,sorting=none]{biblatex}

\usepackage[a4paper, margin=3cm]{geometry}
\usepackage{listings}
\usepackage{graphicx}
\usepackage{verbatim}
\usepackage{stackengine}
\usepackage{subcaption}
\usepackage{siunitx}
\usepackage{xspace}
\usepackage{amsmath}
\usepackage{tablefootnote}
\usepackage{amssymb}
\usepackage[section]{placeins}

\newcommand{\bat}{BAT.jl\xspace}

\newcommand\xrowht[2][0]{\addstackgap[.5\dimexpr#2\relax]{\vphantom{#1}}} 
\usepackage{multirow}
\usepackage[autoload=false]{jlcode}
\begin{document}

\title{BAT.jl --- A Julia-based tool for Bayesian inference}
\author[1]{Oliver Schulz}
\author[2]{Frederik Beaujean}
\author[1]{Allen Caldwell}
\author[3]{Cornelius Grunwald}
\author[1]{Vasyl Hafych}
\author[3]{Kevin Kr{\"o}ninger}
\author[3]{Salvatore La Cagnina}
\author[3]{Lars R{\"o}hrig}
\author[1]{Lolian Shtembari}
\affil[1]{Max Planck Institute for Physics, Munich}
\affil[2]{formerly C2PAP, Excellence Cluster Universe, Ludwig-Maximilian University of Munich}
\affil[3]{TU Dortmund University, Dortmund}
\date{August 2020}

\maketitle

\begin{abstract}
We describe the development of a multi-purpose software for Bayesian statistical inference, \bat, written in the Julia language. The major design considerations and implemented algorithms are summarized here, together with a test suite that ensures the proper functioning of the algorithms. We also give an extended example from the realm of physics that demonstrates the functionalities of \bat.
\end{abstract}

% -----------------------------------------------------------------------
\section{Introduction}
\label{sec:introduction}

The analysis of data with means of statistical methods is a key aspect of scientific research. Depending on the field of research, the type of data and the size of the corresponding data sets can vary strongly, e.g. few event counts obtained in searches for rare radioactive decays, huge samples of astronomical data or images from medical imaging. The common theme connecting these different types of applications is the statistical analysis of the data. One is typically interested in estimating the free parameters of a scientific model given a particular data set, and in comparing two or more models. Bayesian reasoning allows for this in a consistent and easy-to-interpret way. The key element is the equation by Bayes and Laplace, i.e.
\begin{eqnarray}
p(\boldsymbol{\theta} | \mathcal{D},M) = \frac{p(\mathcal{D} | \boldsymbol{\theta}, M) p(\boldsymbol{\theta} | M)}{\int \mathrm{d}\boldsymbol{\theta} \, p(\mathcal{D} | \boldsymbol{\theta}, M) p(\boldsymbol{\theta} | M)} \, ,
\end{eqnarray}
where the term on the left-hand side, $p(\boldsymbol{\theta} | \mathcal{D},M)$, is the posterior probability (density)~\footnote{For better readability, we use the terms probability and probability density synonymously in the following.} for the set of free parameters $\boldsymbol{\theta}$ given a data set $\mathcal{D}$ and assuming a model $M$. It is proportional to the product of the likelihood, $p(\mathcal{D}|\boldsymbol{\theta}, M)$, and the prior knowledge about the parameters, $p(\boldsymbol{\theta}|M)$. The denominator is often referred to as the evidence $Z$; it is the probability to have observed the data $\mathcal{D}$ given the model $M$:
\begin{equation}
    \label{eq:evidence}
    Z = P(\mathcal{D}|M) = \int \mathrm{d}\boldsymbol{\theta} \, p(\mathcal{D} | \boldsymbol{\theta}, M) p(\boldsymbol{\theta} | M) \;\; .
\end{equation}

 The evidence $Z$ is required for model comparison.
 
 Inference about individual parameters can be performed using the multi-dimensional posterior probability or the marginalized probabilities
\begin{eqnarray}
p(\theta_{i} | \mathcal{D}, M) = \int p(\boldsymbol{\theta} | \mathcal{D},M) \, \prod_{i \neq j} \mathrm{d} \theta_{j} \, .
\label{eqn:marginal}
\end{eqnarray}
 We refer to commonly available textbooks for a general introduction to Bayesian inference as well as for the techniques and measures typically used, see e.g. Refs.~\cite{DAgostini2003,Hartigan1983,Jaynes2003,Kendall1994,MacKay2003,Sivia2006}.

In most scientific applications, the model $M$ results in a non-trivial form of the likelihood, such that assumptions that allow using common approximations do not hold( e.g., a Gaussian shape of the likelihood or a linear connection between the predictions and the model parameters). In such cases, it is often necessary to calculate integrals of the type appearing in Eqn.~\ref{eqn:marginal} numerically. Efficient and reliable algorithms are an important aspect of such an evaluation, in particular for models with many parameters, or, more technically, many dimensions of integration. Similar arguments hold for the optimization problem of finding the best-fit parameters associated with the global or marginal modes of the posterior probability. A variety of automated tools are available, usually tailored to the needs of a particular field of research or a class of statistical models, such as STAN~\cite{JSSv076i01}, PYMC~\cite{PYMC}, R~\cite{R2017} or OpenBUGS~\cite{Lunn2000}. An important criterion to choose one tool over the others is its compatibility with the rest of the infrastructure used in a research field, typical data bases or programs used for processing the results obtained.

Due to the lack of such a tool in the field of particle physics, we originally developed the Bayesian Analysis Toolkit (BAT)~\cite{Caldwell:2008fw}, as a C++ library under the open-source LGPL license. It features several numerical algorithms for optimization, integration and marginalization with a strong focus on the use of Markov Chain Monte Carlo (MCMC) algorithms. BAT has been widely used in our field of research and examples of advanced applications in particle physics are global fits of complex models~\cite{Bevan:2014tha,Ghosh:2015wiz,Ciuchini:2014dea,Ciuchini:2013pca,deBlas:2014ula,Agostini:2017jim,Caldwell:2017mqu} and kinematic fitting~\cite{Erdmann:2013rxa}. Over time, BAT-C++ gained traction outside of particle physics as well. It has also been used in many other fields of research; for example in cosmology~\cite{Luongo:2015zgq}, astrophysics~\cite{Ullio:2016kvy}, and nuclear physics~\cite{Rappold:2015una}. The sampling methods implemented in BAT-C++ have also been used to develop more advanced sampling algorithms~\cite{Caldwell:2014eca,Kroeninger:2014bwa}. 

Given the wide range of possible applications, we began to develop a more easily portable version of BAT that does not come with the heavy dependencies on particle-physics software stacks and that also allows for smart parallelization. This development resulted in \bat~\cite{bat.jl}, a completely re-designed BAT implemented in Julia~\cite{DBLP:journals/corr/BezansonEKS14}.

Here, we describe the design, features and numerical performance of the upcoming version 2.0 of \bat. It is is available at \url{https://github.com/bat/BAT.jl/tree/master} under the MIT open-source license~\cite{MITLicense}, and documented at \url{https://bat.github.io/BAT.jl/dev/}. The documentation also includes tutorials that new users can run and modify to quickly familiarize themselves with \bat.

This paper is organized as follows: Section~\ref{sec:design_and_software} describes the considerations that went into the design of the software and the code. Section~\ref{sec:functionality} summarizes the numerical algorithms available in \bat and section~\ref{sec:output} the options provided to output and visualize the numerical results. Tests on the numerical performance of the algorithms is reported on in section~\ref{sec:benchmark} and an extended example demonstrating the strength of \bat is introduced in section~\ref{sec:example}. Section~\ref{sec:summary} provides a summary.

% -----------------------------------------------------------------------
\section{Design considerations and software design}
\label{sec:design_and_software}

\subsection{Design considerations}
\label{sec:design-considerations}

\bat aims to help solve a wide range of complex and computationally demanding problems. The design of the implementation is guided by the requirement to support multi-threaded and distributed code and offers a choice of sampling, optimization and integration algorithms. At the same time, we want to offer a user-facing interface that makes it easy to quickly solve comparatively simple problems, while offering the direct access to lower-level functionality and tuning parameters that an expert may need to solve very hard problems. Finally, we wanted to make it very easy for the user to interact with and visualize results of \bat's algorithms.

We chose to implement \bat in Julia due to Julia's unique advantages for statistical and other numerical applications that require high numerical performance and easy composability of different algorithms.

Julia allows for writing code in an easy fashion, similar to Python, but at the same time enables that code to run with very high performance, like code written in C, C++ or FORTRAN. In addition, Julia is one of the few languages based on multiple-dispatch---this solves the expression problem~\cite{Zenger:52625} and therefore results in a level of code-composability superior to object-oriented (i.e. single-dispatch) languages. This is complemented by Julia's state-of-the-art package manager that makes if very easy for the user to install third-party packages.

Julia also enables automatic differentiation of almost arbitrary code via both multiple-dispatch~\cite{RevelsLubinPapamarkou2016} and via it's LISP-like meta-programming capabilities~\cite{Innes:2018zygote}. This makes it possible to use gradient-based algorithms like HMC-sampling~\cite{DUANE1987216, doi:10.1201/b10905-7, 2017arXiv170102434B} and L-BFGS optimization~\cite{Mogensen2018} with automatic differentiation, so the user is not required to provide a hand-written gradient for likelihood and prior densities. Julia code can be run on both CPUs and GPUs~\cite{besard2018juliagpu}. The language also offers first-class support for writing multi-threaded and distributed code. These features significantly lower the effort required when tackling problems that require highly efficient code and massive computational resources.

Julia also has very good support for interfacing with code written in C, FORTRAN, C++, PYTHON and many other languages. So while \bat itself is written in Julia, the user can easily access likelihood functions written in another language, typically with minimal or no impact on performance. This is important when the likelihood functions include complex existing (e.g. in physical or biological) models.

\bat is designed to integrate well with related packages in the Julia software ecosystem. To further improve this integration and code-reuse, we have released functionalities that may be also useful outside of \bat's main scope as separate packages, e.g. ArraysOfArray.jl, ValueShapes.jl and EmpiricalDistributions.jl. As such, \bat is modular, and we aim to improve this modularity in future releases.

\subsection{Software design}
\label{sec:software-design}

The software model of \bat is centered on positive-definite densities. These may be normalized (and can then be viewed as probabilities) or not: likelihoods, priors and posteriors are all expressed as densities (represented by the type \jlinl{AbstractDensity}). \bat automatically converts user-provided density-like objects like log-likelihood functions, distributions and histograms to subtypes of \jlinl{AbstractDensity}.

Julia's unique advantages as a multi-dispatch programming language allow us to provide a very compact user-facing API that still makes it possible to build complex statistical analysis chains based on fundamental operations like sampling, optimization and integration.

To operate on densities, \bat offers functions like \jlinl{bat_sample}, \jlinl{bat_findmode} and \jlinl{bat_integrate}. These can be combined in a very flexible and intuitive fashion: \jlinl{bat_sample} will automatically try to sample from prior densities via iid (independent and identically distributed) sampling, from posterior densities via MCMC and from existing samples themselves via resampling. \jlinl{bat_findmode} and \jlinl{bat_integrate} will automatically sample, use optimization algorithms or analyse existing samples, depending on the given density. 

\bat has a unified mechanism to manage default behavior and algorithmic choices. The function \jlinl{bat_default} lets the user query which algorithm with which settings would be used for a given task. \bat also records the choice of algorithms and their configuration (whether explicit or implicit) in it's results. In general, \bat will always try to choose an appropriate default strategy for a given task, but will let the user override default choices for algorithms and configuration or tuning parameters.

To take advantage of the parallel architecture of modern computer systems, \bat uses Julia's advanced multithreading scheduler to parallelize operations automatically where possible. For example, MCMC chains automatically run on separate threads while the user can still use multi-threading within the implementation of the likelihood function to further load out the processors of the system, without over-subscription. MCMC sampling and integration can also be run on multiple remote hosts, using Julia's support for compute clusters. MPI message transport can be used when available, but a plain TCP/IP network is sufficient.

We take great care to ensure that results are reproducible, independent of the possibly multi-threaded and distributed computation strategy. BAT uses a hierarchical scheme to partition and distribute counter-based random number generators (RNGs). By default, BAT uses the Philox RNG~\cite{Salmon2011} to generate random numbers. We automatically partition this counter space (using a safe upper limit for the possible amount of RNG generation in each separate computation). Each MCMC chain, and even each step of each MCMC chain, effectively uses it's own independent RNG - no matter which resources that step is scheduled to be computed on. If computations are hierarchical, each partition of an RNG counter space can be partitioned again and again, following the graph of the computation. The counter space of generators like Philox typically consists of two or four 64-bit numbers. So even nested parallel computations, each with an ample reserve of random numbers, will not run out of counter space.

% -----------------------------------------------------------------------
\section{Numerical algorithms}
\label{sec:functionality}

Several algorithms for marginalization, integration and optimization are implemented in \bat, giving it a toolbox character that also allows for the future inclusion of further methods, algorithms and software packages. The central algorithms available in \bat are summarized in the following. We do not go into detail on additional minor functionalities, like simple evaluation of the probability distribution on a grid for a small number of dimensions and the usage of quasirandom sequences.

\subsection{Sampling algorithms}

BAT.jl currently provides a choice of two main MCMC sampling algorithms to the user, Metropolis-Hastings (MH) and Hamiltonian Monte Carlo (HMC). Different algorithms are more or less suited for different target densities - for example, HMC sampling cannot be used if the target is not differentiable.

\subsubsection{Metropolis-Hastings}
The Metropolis-Hastings algorithm~\cite{Metropolis:1953am} is the original MCMC algorithm to produce a random set of numbers $\theta$ or vectors $\boldsymbol{\theta}$ that have the properties of a Markov chain and that converge towards a target distribution. In Bayesian analysis, the limiting distribution of this set $\pi(\boldsymbol{\theta})$ will be the posterior probability density $p(\boldsymbol{\theta} | M)$. The samples are generated as follows: starting from a state $\boldsymbol{\theta}_{i}$ at iteration $i$, a new state $\boldsymbol{\theta'}$ is proposed according to a (often symmetric) proposal distribution $g(\boldsymbol{\theta'} | \boldsymbol{\theta})$. The proposal is accepted with a probability

\begin{equation}
P_{\rm{accept}} = {\rm min}\left(1, \frac{\pi(\boldsymbol{\theta'})}{\pi(\boldsymbol{\theta}_{i})}\ \frac{g(\boldsymbol{\theta}_{i} | \boldsymbol{\theta'})}{g(\boldsymbol{\theta'} | \boldsymbol{\theta}_{i})}\right)
\end{equation}

\noindent resulting in $\boldsymbol{\theta}_{i+1} = \boldsymbol{\theta'}$, or $\boldsymbol{\theta}_{i+1} = \boldsymbol{\theta}_{i}$ if the proposal is rejected. We run several Markov chains in parallel and repeatedly test for convergence during a burn-in phase (see \ref{sec:burnin}).

By default, \bat uses a multivariate Student's t distribution as the proposal distribution. The scale and correlation of the proposal is adapted automatically in order to efficiently generate samples from essentially any smooth, unimodal distribution. Another important characteristic of Markov chains is the acceptance rate $\alpha$, the ratio of accepted proposal points to the total number of samples in the chain. For any given target and proposal distribution there is an optimal $\alpha$ that will allow the best exploration and performance of the chain.

In order to achieve a desired acceptance ratio the proposal distribution is tuned to adapt it to the target. After each tuning cycle (see \ref{sec:burnin}), the covariance matrix of the proposal function, $\boldsymbol{\Sigma}$, is updated based on the sample covariance of the last iterations and it is then multiplied with a scale factor $c$ that governs the range of the proposal. $c$ is tuned to force the acceptance rate to lie in a region of $\alpha_{min} \leq \alpha \leq \alpha_{max}$ and is restricted to the region $c_{min}  \leq c \leq c_{max}$. The adjustment of the scale factor is descried in Algorithm 1 of \cite{FrederikBeaujean}. The default values in \bat for the acceptance rate and scale factor ranges are $\alpha_{min} = 0.15$, $\alpha_{max} = 0.35$ \cite{RGG97} and $c_{min} = 10^{-4}$, $c_{max} = 100$ respectively.

\subsubsection{Hamiltonian Monte Carlo}
One of the most sophisticated MCMC sampling methods is Hamiltonian Monte Carlo (HMC) \cite{DUANE1987216, doi:10.1201/b10905-7, 2017arXiv170102434B}. By using a proposal function that is adjusted to the shape of the target distribution, HMC algorithms can yield higher acceptance rates and less correlated samples than other sampling algorithms based on random walks, thus reducing the number of samples required to fully explore the target distribution.

In HMC, the $D$-dimensional parameter space is expanded to $2D$ dimensions by introducing so-called momenta $\vec{p}$ as hyperparameters, moving from the original phase space to the canonical phase space $ \vec{q} \rightarrow (\vec{q}, \vec{p})$. In order to conform to standard notation when discussing HMC, we here use $\vec{q}$ to represent the parameters of the model in place of $\boldsymbol{\theta}$. 

In the HMC formalism, the target distribution $\pi(\vec{q})$ is lifted to the canonical phase space using a joint probability distribution
\begin{equation}
    \pi(\vec{q}, \vec{p}) = \pi(\vec{p}|\vec{q}) \pi(\vec{q}) = \mathrm{e}^{-H(\vec{q},\vec{p})}\,, \label{HMCprob}
\end{equation}
where the probability distribution of the momenta $\pi(\vec{p}|\vec{q})$ is chosen to be conditional.
The last equality in Eq.~\eqref{HMCprob} comes from defining the so-called Hamiltonian as
\begin{equation}
    H(\vec{q},\vec{p}) = -\log \pi(\vec{q}, \vec{p}) = -\log\pi(\vec{p}|\vec{q}) -\log\pi(\vec{q})\,.
\end{equation}

The differential equations 
\begin{equation}
    \frac{dq_i}{dt} = \frac{\partial H}{\partial p_i}, \quad \frac{dp_i}{dt} = -\frac{\partial H}{\partial q_i}\,,
\end{equation}
are well known from classical mechanics and referred to as the Hamilton's equations of motion. Solving the equations of motion for a certain time $T$ allows moving along trajectories $\phi$ and gives a transition in the canonical phase space
\begin{equation}
    (\vec{q}, \vec{p}) \rightarrow \phi_T(\vec{q}, \vec{p}) = (\vec{q}^*, \vec{p}^*)\,,
\end{equation}
resulting in the new point $(\vec{q}^*, \vec{p}^*)$.
By marginalizing over the momenta $\vec{p}$, we obtain a new proposal point $\vec{q}^*$ in the original parameter space. This proposal is then either accepted as a new sampling point or rejected by calculating an acceptance ratio, similar to the MH algorithm. Since the proposal points are generated using information of the target distribution, their acceptance rates are higher than samples using non-problem-specific proposal distributions.

Since HMC requires gradient information and introduces multiple hyperparameters (such as momenta and integration times) into the sampling process, performing Bayesian analyses with HMC samplers is usually not as straight-forward as using the MH algorithm as it requires additional computational steps such as the numerical integration of the equations of motions and the selection and tuning of the hyperparameters. \bat uses the AdvancedHMC.jl package~\cite{ge2018t} for the single HMC sampling steps. AdvancedHMC.jl provides several flavours of HMC, including multiple versions of the No-U-Turn Sampler (NUTS)~\cite{JMLR:v15:hoffman14a}. Higher level operations and the burn-in process are handled by \bat itself, like for MH sampling. Due to the efficient support of automatic differentiation in Julia, e.g. through the package ForwardDiff.jl \cite{RevelsLubinPapamarkou2016}, the gradient of the target, required for HMC, can often be derived automatically. This makes it quite easy to use HMC within BAT.

\subsubsection{MCMC burn-in process} 
\label{sec:burnin}

 Different MCMC samling algorithms have different tuning parameters, e.g. the scale and shape of the proposal function for MH. But a common requirement for the generation of samples that faithfully follow the target density is a suitable burn-in process: Starting with an initial sample, each MCMC chain must be allowed to run until is has converged to it's stationary distribution. Several MCMC chains must be compared to ensure that they share the same stationary distribution and are not, for example, limited to different modes of the posterior.

\bat will by default use four MCMC chains, which are iterated in parallel on multiple threads (and in the future, also on multiple compute nodes). We initialize each MCMC chain with a random sample drawn from the prior, and we require that efficient sampling is possible for all priors. Typically, priors will be composed from common distributions provided by the Julia package Distributions.jl, which supports iid sampling for all of it's distributions.

Once the MCMC chains are initialized, burn-in, MCMC tuning and convergence testing are performed in cycles. The user specifies the desired number of samples after burn-in, the length of each tuning/burn-in-cycle is by default 10\% of desired number of final samples. During each cycle, each MCMC chain is iterated and tuning parameters are adjusted in an algorithm-specific fashion. At the end of each cycle, we check for convergence of all MCMC chains. Tuning and burn-in are complete when all chains are tuned (according to algorithm-specific criteria) and have converged (see below). MCMC samples produced until the point are discarded by default, then chains are run for the desired number of steps (the user can also set limits like maximum wall-clock time) without further modification of tuning parameters. If tuning and convergence are not successful within a (user adjustable) maximum number of cycles, the user has the option between receiving a warning message or the  sampling to terminate with an error exception.

\subsubsection{Convergence Tests} 
\label{sec:convergence}
In order to determine if the Markov chains have converged and the burn-in phase can stop, we adopt the Gelman-Rubin convergence test~\cite{Gelman-Rubin:1992gr} and the Brooks-Gelman test~\cite{Brooks-Gelman:1998bg} (our default). 

We consider first a single parameter $\theta$ and running $N$ chains in parallel, where each chain produces $M$ samples: ${\theta_{1i}, ..., \theta_{Ni}}$ (where $i = 1,...,M$). The Gelman-Rubin test relies on two estimators of the variance of $\theta$: the within-chain variance estimate, 
    \begin{equation}
    \label{eq:in-chain-variance}
    W = \sum_{i = 1}^{M} \sum_{j = 1}^{N} \frac{(\theta_{ij} - \bar{\theta_i})^2}{M(N-1)} ;
    \end{equation}
and the pooled variance estimate
    \begin{equation}
    \label{eq:between-chain-variance}
    \hat{V} = \frac{(N-1)W}{N} + \sum_{i=1}^{M} \frac{(\bar{\theta_i} - \bar{\theta})^2}{M-1} \, ,
    \end{equation}
where $\bar{\theta_i}$ is the $i$-th chain mean and $\bar{\theta}$ is the overall mean. Using these estimators we construct the potential scale reduction factor (PSRF) denoted by $\hat{R}$,

\begin{equation}
\label{eq:PSRF}
\hat{R} = \frac{\hat{V}}{W} \, .
\end{equation}

Since the $N$ chains are randomly initiated from an over-dispersed initial distribution, within a finite number of samples per chain, $\hat{V}$ overestimates the target variance while $W$ underestimates it. This implies that $\hat{R}$ will have a value larger than 1 and the degree of convergence of the chains is measured by the closeness of $\hat{R}$ to the value 1. 

So we construct the  multivariate PSRF (MPSRF) denoted by $\hat{R_p}$,

\begin{equation}
\label{eq:MPSRF}
\hat{R_p} = %\ma\theta_{\{a\}} \frac{a^T \hat{V^*} a}{a^T W^* a} = 
\frac{N-1}{N} + \left( \frac{M+1}{M} \right) \Lambda_1
\end{equation}

\noindent where the variance estimates are 

\begin{equation}
\label{eq:in-chain-variance-multivariate}
W^* = \sum_{i = 1}^{M} \sum_{j = 1}^{N} \frac{(\boldsymbol{\theta}_{ij} - \boldsymbol{\bar{\theta}}_i)(\boldsymbol{\theta}_{ij} - \boldsymbol{\bar{\theta}}_i)^T}{M(N-1)}
\end{equation}

\begin{equation}
\label{eq:between-chain-variance-multivariate_B}
\frac{B^*}{N} = \sum_{i=1}^{M} \frac{(\boldsymbol{\bar{\theta_i}} - \boldsymbol{\bar{\theta}})^2}{M-1}
\end{equation}

\begin{equation}
\label{eq:between-chain-variance-multivariate_V}
\hat{V}^* = \frac{(N-1)W}{N} + \frac{B^*}{N}
\end{equation}

\noindent and $\Lambda_1$ is the largest eigenvalue of the matrix $\frac{{W^*}^{-1}B^*}{N}$. The default cut-off we use to declare convergence in the burn-in phase is $\hat{R}, \hat{R}_p \leq 1.1$.\\

\subsubsection{Effective Sample Size} 
\label{sec:ESS}

A drawback of MCMC is that the samples we obtain are correlated. \bat provides an effective sample size (ESS) estimator to calculate what number of iid samples would be equivalent to $N$ given MCMC samples, in respect to the variance of sample-mean estimates. It is also a valuable indicator on whether a sufficient number of MCMC samples has been produced.

The effective sample size is estimated as:

\begin{equation}\label{ESS}
    \mathrm{ESS} = \frac{N}{\hat{\tau}}
\end{equation}

\noindent where $\hat{\tau}$ is the integrated autocorrelation time. $\hat{\tau}$ is estimated from the normalized autocorrelation function $\hat{\rho}(\tau)$:

\begin{equation}\label{integrated_autocorrelation_time}
    \hat{\tau}_k = 1 + 2 \sum_{\tau = 1}^{\infty} \hat{\rho}_k(\tau)
\end{equation}

\begin{equation}\label{normalized_autocorrelation_function}
    \hat{\rho}_k(\tau) = \frac{\hat{c}_k(\tau)}{\hat{c}_k(0)}
\end{equation}

\begin{equation}\label{c_autocorrelation_function}
    \hat{c}_k(\tau) = \frac{1}{N - \tau} \sum_{n = 1}^{N - \tau} \left( \theta_{k,i} - \hat{\theta}_k \right) \left( \theta_{k,i+\tau} - \hat{\theta}_k \right)
\end{equation}

\noindent where $k$ refers dimension index of the multivariate sample $\boldsymbol{\theta}_i = \{ \theta_{1, i}, ...,\theta_{D, i} \}$. Here, all samples for a given parameter $k$ are used, independently of whether multiple chains have been run.  The first index refers now to the variable under discussion (as opposed to the chain number, above).  These quantities allow us to calculate an effective sample size for each dimension $\mathrm{ESS}_k = \frac{N}{\hat{\tau}_k}$.

When evaluating Eq.\ref{integrated_autocorrelation_time} we can't, in practice, actually sum over all lags $\tau$: while $\hat{c}_k(\tau)$ theoretically decays to zero for high lags $\tau$, in practice it exhibits a noisy behavior that makes the sum over $\hat{c}_k(\tau)$ unstable. So we need to truncate the sum using a heuristic cut-off. The default cut-off in \bat is Geyer's initial monotone sequence estimator \cite{Geyer:1992}, optionally Sokal's method \cite{MadrasSokal:1988} can be chosen.

\subsection{Algorithms for point estimates}
The global mode of a posterior distribution is often a quantity of interest. While the MCMC sample with the largest value of the target density may come close to the true mode, it is sometimes not as close as required. It is, however, an ideal starting point for a local optimization algorithm than can then further refine the mode estimation. \bat offers automatic mode-estimation refinement using the Nelder–Mead~\cite{Nelder1965ASM} and LBFGS~\cite{LBFGS1998} optimization algorithms, by building on the Optim.jl~\cite{Mogensen2018} package. When using LBFS, a gradient of the posterior distribution is required. Again, we utilize the Julia automatic-differentiation package ecosystem to automatically compute that gradient.

Another quantity that is often computed from samples is a marginal mode. To construct marginals, a binning of the samples is performed. The optimal number of bins can be determined by using Square-root choice, Sturges' formula, Rice Rule, Scott's normal reference rule, or Freedman–Diaconis rule. The latter is the default.

BAT.jl also provides functionality to estimate other quantities such as the median, the mean, quantiles and standard deviations, and to propagate errors on a fit function.

\subsection{Integration algorithms}

\subsubsection{Evidence Estimation using AHMI}
In many applications, it is desirable or even necessary to compute the evidence or marginal likelihood $Z$ (see Eq.~\ref{eq:evidence}). An example for the use of $Z$ is the calculation of a Bayes factor for the comparison of two models $M_A$ and $M_B$:
$$
{\rm BF} \equiv \frac{p(\mathcal{D}| M_A)}{p(\mathcal{D}| M_{B})} = \frac{Z_A}{Z_{B}} \; .
$$
\bat includes the Adaptive Harmonic Mean Integration (AHMI) algorithm~\cite{ahmi} to compute $Z$ given the samples $\{ \boldsymbol{\theta} \}$. 

AHMI can integrate samples from any sampling algorithm as long as the samples come in the form of \jlinl{BAT.DensitySampleVector}. It's use of hyper-rectangles, however, limits the applicability to a moderate number of dimensions ($\approx 20$ in the case of a multivariate normal distribution).

\subsubsection{Evidence calculation using an interface to CUBA}

In addition to integration via AHMI, BAT offers evidence calculation using the Cuba~\cite{Hahn:2004cuba} integration library. Cuba implements multiple integration algorithms that cover a range of (Monte-Carlo and deterministic) importance sampling, stratified sampling and adaptive subdivision integration strategies. These will typically not scale to high-dimensional spaces, but can provide quick and robust results for low-dimensional problems.

% -----------------------------------------------------------------------
\section{Output and visualization of results}
\label{sec:output}

The results of running the numerical algorithms in \bat are presented in text and also in graphical form. In addition, user-defined interfaces can be written to bring the results into any other format.

\subsection{Graphical summary of the results}
As a key element of all statistical analyses is the graphical representation of outcomes, \bat includes functionalities to create visualizations of the analyses results in a user-friendly way.
By providing a collection of plot recipes to be used with the Plots.jl\footnote{\url{https://github.com/JuliaPlots/Plots.jl}} package, several plotting styles for 1D and 2D representations of (marginalized) distributions of samples and priors are available through simple commands. Properties of the distributions, such as highest density regions or point estimates like mean and mode values, can be automatically highlighted in the plots. Further recipes to visualize the results of common applications, such as function fitting, are provided.
While the plot recipes provide convenient default options, the details of the plotting styles can be quickly modified and customized. Since all information about the posterior samples and the priors are available to the user, completely custom visualizations are of course also possible. Examples of plots created with the included plot recipes are shown in section~\ref{sec:example}.

\subsection{Written summary of the results}

\bat can display a written summary containing information about the sampling process and the results of the parameter estimation (i.e. mean, median and quantiles for each parameter). Additional functions provide access to specific results or additional information.

Extending and customizing the default outputs or implementing custom output formats is possible.

\subsection{File I/O}

To make it possible to preserve the results of the (often computationally expensive) MCMC sampling process, \bat provides explicit functions to store MCMC sample variates, weights and log-density values in HDF5 files, and can read them again at a later time. Samples can also be easily written to ASCII/CSV-files using standard Julia functionalities.

% -----------------------------------------------------------------------
\section{Numerical test suite}
\label{sec:benchmark}

A test suite to evaluate the numerical performance of the sampling algorithms is included in \bat, and must be passed before each release of a new version. Samples are MCMC-generated from, and then compared to, a set of test distributions. A list there distributions is given in Tab.~\ref{tab:performance_functions_2D}. We compare the mean values, variances, and the global modes of the samples with those of the test distributions.  We also calculate the p-values of Kolmogorov-Smirnov (KS) tests for each parameter, by comparing the marginal distributions from the sampling algorithm with marginal distributions from samples generated by iid sampling. Small p-values lead to further investigations to ensure that the sampling algorithm is functioning properly.

Additionally, the integral of the target distributions is calculated from the samples using AHMI. Since AHMI relies on an accurate sampling of the target distribution, the AHMI integral value provides a very sensitive test of the sampling algorithm. 

\begin{table}
    \centering
	\caption{Listing of the analytical form of two dimensional test functions used for performance testing.}
	\label{tab:performance_functions_2D}
	\resizebox{0.7\textwidth}{!}{
	\begin{tabular}{c|c} \hline\hline
			\textbf{name} & \textbf{function} \\ \hline  \xrowht{20pt}
			normal 				& $f({x})={\frac {\exp \left(-{\frac {1}{2}}({ {x} }-{ {\mu }})^{\mathrm {T} }{{\Sigma }}^{-1}({ {x} }-{{\mu }})\right)}{\sqrt {(2\pi )^{k}|{{\Sigma }}|}}}$\\ \xrowht{20pt}
			multi cauchy   		& $f\left( \lambda \right )  = \prod_{i=1}^2 \frac{1}{2} \left[ \mathrm{Cauchy}\left( \lambda_i \mid  \mu, \sigma \right) + \mathrm{Cauchy} \left( \lambda_i \mid - \mu, \sigma \right) \right]$ \\ \xrowht{20pt}																	
																			
			funnel  		& $f\left(\lambda \right)  =  \mathcal{N}\left( \lambda_1 \mid 0, a^2 \right) \prod_{i=2}^n  \mathcal{N}\left( \lambda_i \mid 0, \exp\left( 2b\lambda_1 \right) \right)$ \\ \hline\hline %\xrowht{10pt}																	
\end{tabular}

%gaussian shell   	& $f(\lambda| \vec{c},r,\omega)=\frac{1}{\sqrt{2\pi\omega^2}}\exp\left(-\frac{(\vert \lambda-\vec{c}\vert-r)^2}{2\omega^2}\right)$ \\ \xrowht{20pt}		
	}
\end{table}

\begin{figure}
	\centering
	\includegraphics[width=0.8\textwidth]{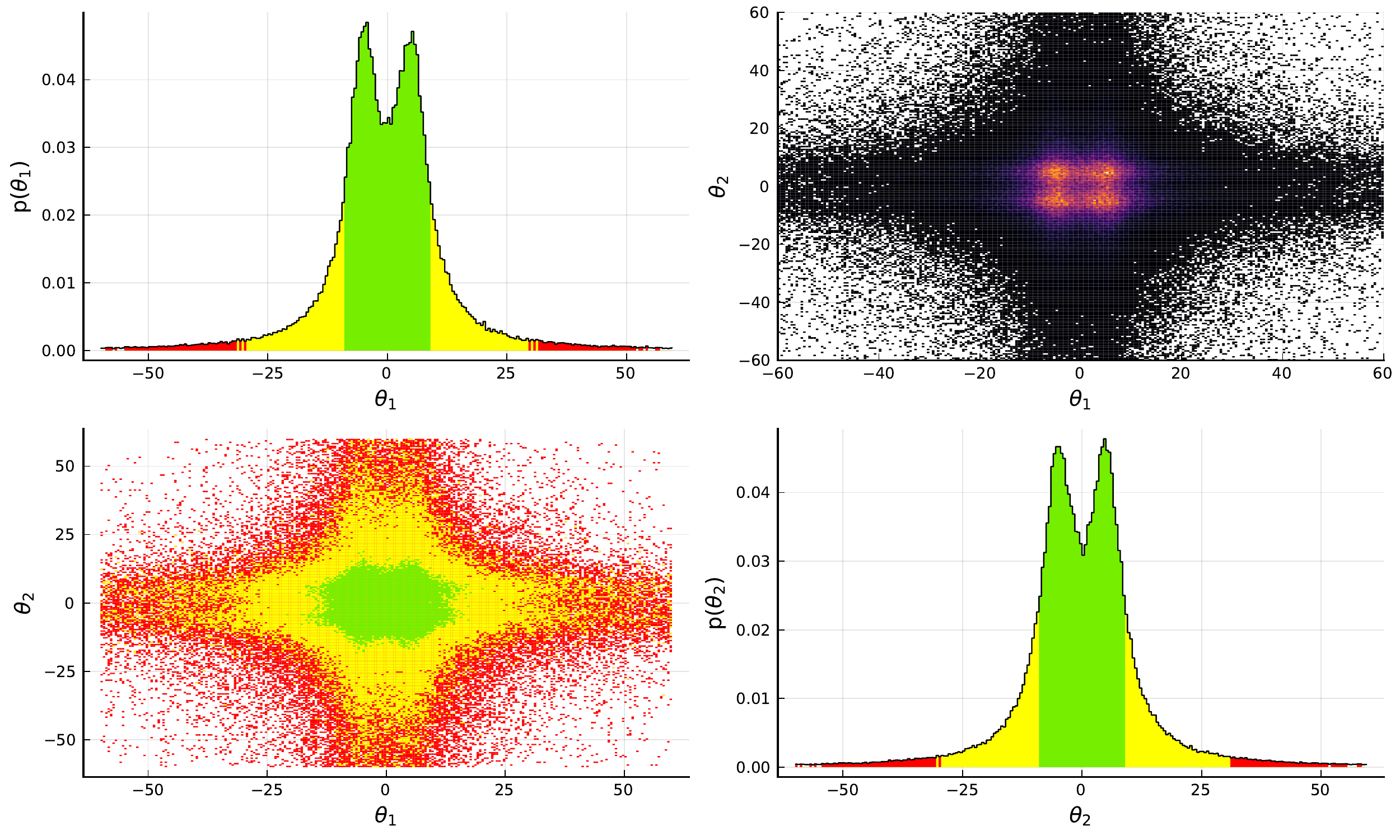}	
	\caption{BAT default plots for the multi-modal Cauchy distribution. The plots on the upper left and lower right show the marginalized distribution for each dimension. The other two plots show the full 2D distribution with the lower left plot focusing on illustrating probability intervals and the upper right one the general shape of the 2D sample. The dashed line indicates the global mode of the sample whilst the green, yellow and red colored samples are defined by the $68.57\%$, $99.5\%$ and $99.7\%$ quantiles. }
	\label{fig:2d_default_cauchy}
\end{figure}

\begin{figure}
	\centering
	\includegraphics[width=0.8\textwidth]{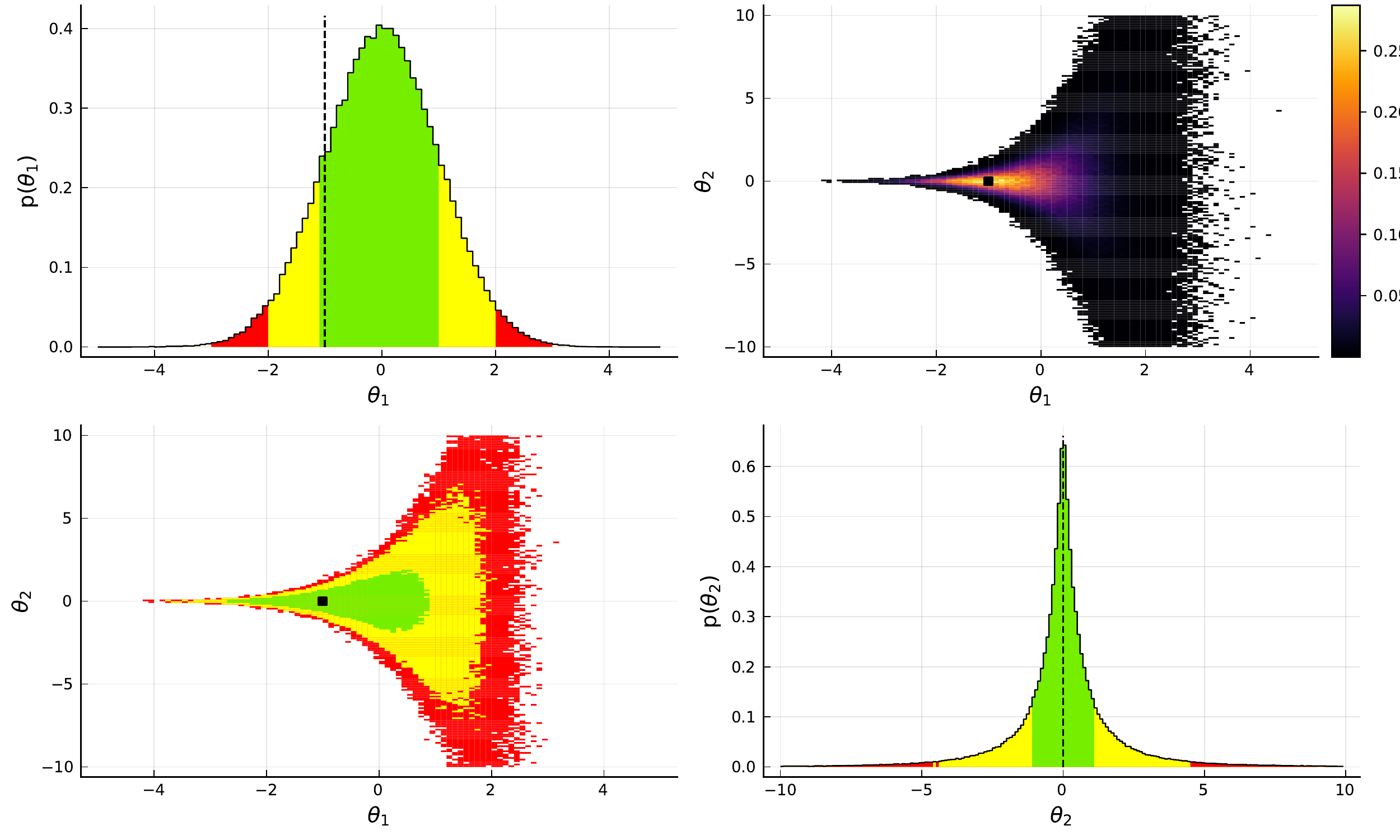}	
	\caption{BAT default plot for the funnel distribution. The plots on the upper left and lower right show the marginalized distribution for each dimension. The other two plots show the full 2D distribution with the lower left plot focusing on illustrating probability intervals and the upper right one the general shape of the 2D sample. The dashed line indicates the mode of the sample whilst the green, yellow and red colored areas represent the $68.57\%$, $99.5\%$ and $99.7\%$ intervals of the sample respectively. Both distributions are normalized to unity. }
	\label{fig:2d_default_funnel}
\end{figure}

\begin{figure}
	\centering
	\begin{subfigure}[b]{0.48\textwidth}
		\includegraphics[width=\textwidth]{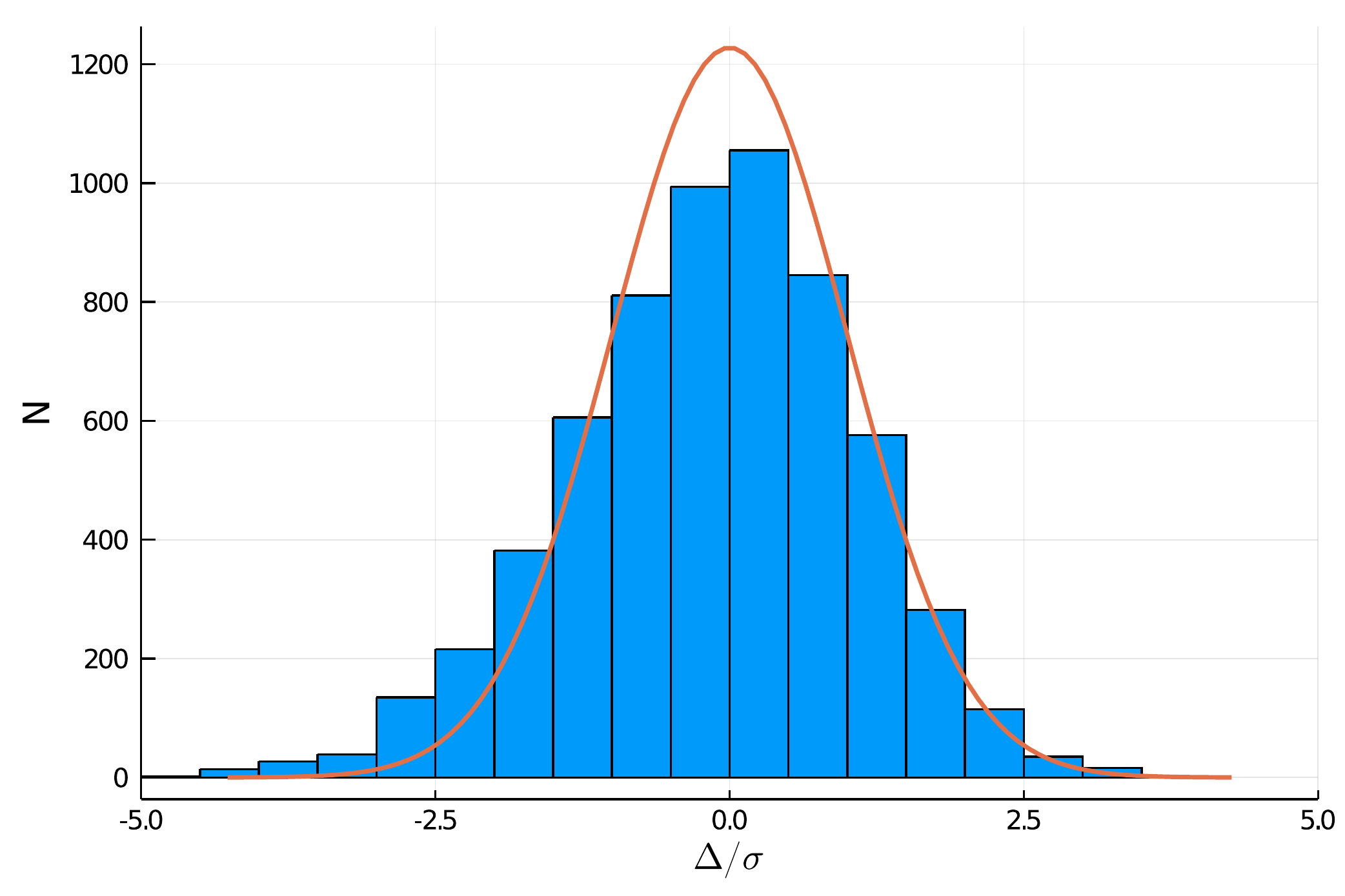}	
	\end{subfigure}
	\hfill	
	\begin{subfigure}[b]{0.48\textwidth}
		\includegraphics[width=\textwidth]{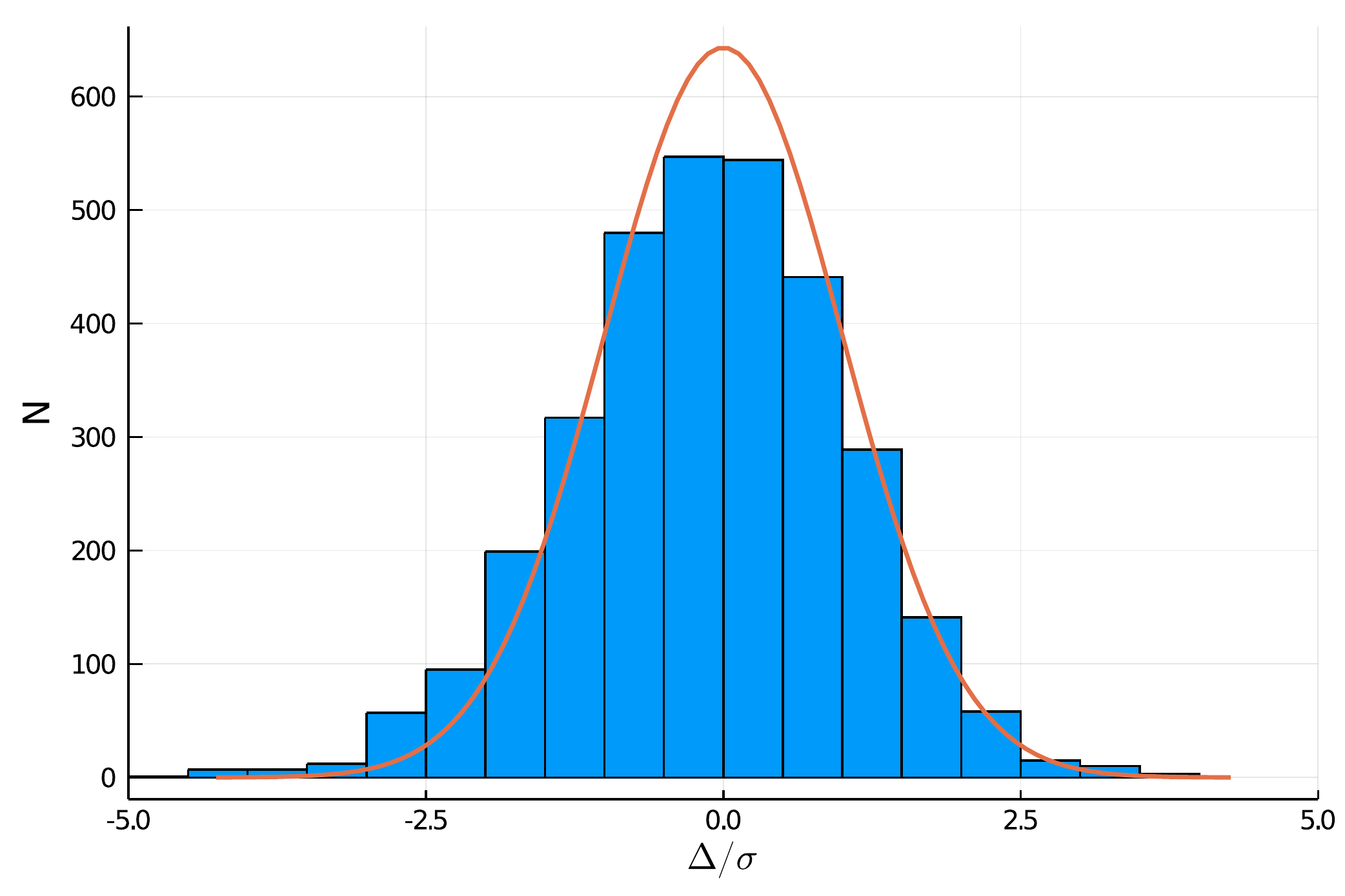}	
	\end{subfigure}
	\caption{The pull plot of the difference between the analytical function (red curve) and the distribution of the samples (blue histogram) for the multi modal Cauchy (left) and funnel distribution (right).}
	\label{fig:2d_cauchy_funnel_pull}
\end{figure}

As an example, Fig.~\ref{fig:2d_default_cauchy} and Fig.~\ref{fig:2d_default_funnel} show the distributions of the samples generated for a multi-modal Cauchy and for the funnel distribution, respectively. 

Assuming iid sampling and a large number of samples, the differences, in units of standard deviations, between the observed distributions and the true distributions are expected to follow a unit normal distribution. This should also be the case if the number of MCMC samples is large enough. For our tests, we compare the expectations in intervals (bins) of the function arguments.  The standard deviation is estimated for each bin as the square root of the expected number of entries from the test function.
For each bin with an expectation larger than ten, the observed number of entries is divided by that standard deviation. 
A histogram of these values, also referred to as pull plot, can be seen in Fig.~\ref{fig:2d_cauchy_funnel_pull}.  It is compatible with expectations.

Table~\ref{tab:perfomance_results_2D} summarizes the expected and observed mean values, variances and global modes for the different two-dimensional test functions, together with the corresponding KS test p-values and AHMI integral values. Very good agreement is observed in all distributions with a maximal deviation of $4 \%$ in the mode, $4 \%$ in the variance.  The AHMI integrals are all very close to the true values, and are typically within the reported uncertainty.  The smallest KS test p-value of $0.023$.  We note that the ESS defined in Section~\ref{sec:ESS} is used in calculating the p-value for the KS test. For the Cauchy distributions, the p-values close to $1$  indicate that the ESS values may be underestimated.  We have noticed that this can occur when the samples become highly correlated.

\begin{table}
	\centering
	\caption{The performance test results for two dimensional functions.}
	\label{tab:perfomance_results_2D}
	\resizebox{\textwidth}{!}{
	%\begin{tabular}{cccc}
%\hline\hline
%\textbf{name} & \textbf{normal} & \textbf{multi cauchy} & \textbf{funnel} \\\hline
%mode target & [15.0, 10.0] & [5.0, 5.0] & [-1.0, 0.0] \\
%mode test & [15.002, 9.998] & [4.796, 4.795] & [-1.003, -0.0] \\
%mode diff (abs) & [-0.002, 0.002] & [0.204, 0.205] & [0.003, 0.0] \\
%mode diff (rel) & [-0.0, 0.0] & [0.041, 0.041] & [-0.003, -] \\
%mean target & [15.0, 10.0] & [0.0, 0.0] & [0.0, 0.0] \\
%mean test & [14.999, 9.995] & [4.336, -0.519] & [0.001, -0.011] \\
%mean diff (abs) & [0.001, 0.005] & [-4.336, 0.519] & [-0.001, 0.011] \\
%mean diff (rel) & [0.0, 0.0] & [-, -] & [-, -] \\
%var target & [2.25, 6.25] & [-, -] & [1.0, 7.407] \\
%var test & [2.248, 6.24] & [21999.687, 4870.124] & [0.993, 6.9] \\
%var diff (abs) & [0.002, 0.01] & [-, -] & [0.007, 0.507] \\
%var diff (rel) & [0.001, 0.002] & [-, -] & [0.007, 0.068] \\
%KS test p-value & 0.955 & 1.0 & 0.949 \\
%AHMI integral & 1.001 & 0.999 & 1.001 \\\hline\hline
%\end{tabular}

\begin{tabular}{c|c|c|c|c|c|c}
\hline\hline
\textbf{name} 	& \multicolumn{2}{|c|}{\textbf{normal}}		& \multicolumn{2}{|c|}{\textbf{multi cauchy}}  		& \multicolumn{2}{|c}{\textbf{funnel}} \\ \hline
				& \textbf{target}	&	\textbf{test}				& \textbf{target}	& \textbf{test} 				& \textbf{target}	& \textbf{test} \\\hline
mode 		 	& [15, 10] & [14.999, 10.002] 	& [5, 5] & [4.793, 4.802]	& [-1, 0] & [-1.001, 0.001] \\
%mode diff (abs) & [0.001, -0.002] 		& [0.207, 0.198] 		 		& [0.001, -0.001] \\
%mode diff (rel) & [0.0, -0.0] 			& [0.041, 0.04] 		 		& [-0.001, -] \\
mean target 	& [15, 10] & [15.002, 10.001]	& [0, 0]  & [-1.352, 0.375]		 		& [0, 0] & [-0.008, -0.005] \\
%mean diff (abs) & [-0.002, -0.001] 		& [1.352, -0.375] 		 		& [0.008, 0.005] \\
%mean diff (rel) & [-0.0, -0.0] 			& [-, -] 			 		& [-, -] \\
var  		& [2.25, 6.25] 	& [2.266, 6.232] 		& [-, -] & [13894, 7442]  		 		& [1, 7.407] & [1.01, 7.085] \\
%var diff (abs) 	& [-0.016, 0.018] 		& [-, -] 			 		& [-0.01, 0.322] \\
%var diff (rel) 	& [-0.007, 0.003] 		& [-, -] 			 		& [-0.01, 0.043] \\
AHMI integral 	& $1.000 \pm 0.003$ & 1 & $1.001 \pm 0.003$ & 1 & $ 1.003 \pm 0.002$ & 1\\
KS test p-value & \multicolumn{2}{|c|}{[0.182, 0.922]}		& \multicolumn{2}{|c|}{[1.0, 1.0]} 			 		& \multicolumn{2}{|c}{[0.023, 0.325]} \\\hline\hline
\end{tabular}

	}
\end{table}

The tests in higher dimensions are performed using the same functions as for the 2D testing.
For these cases, 4 chains each with $2 \cdot 10^5$ samples are generated.

\begin{figure}
	\centering
	\includegraphics[width=0.8\textwidth]{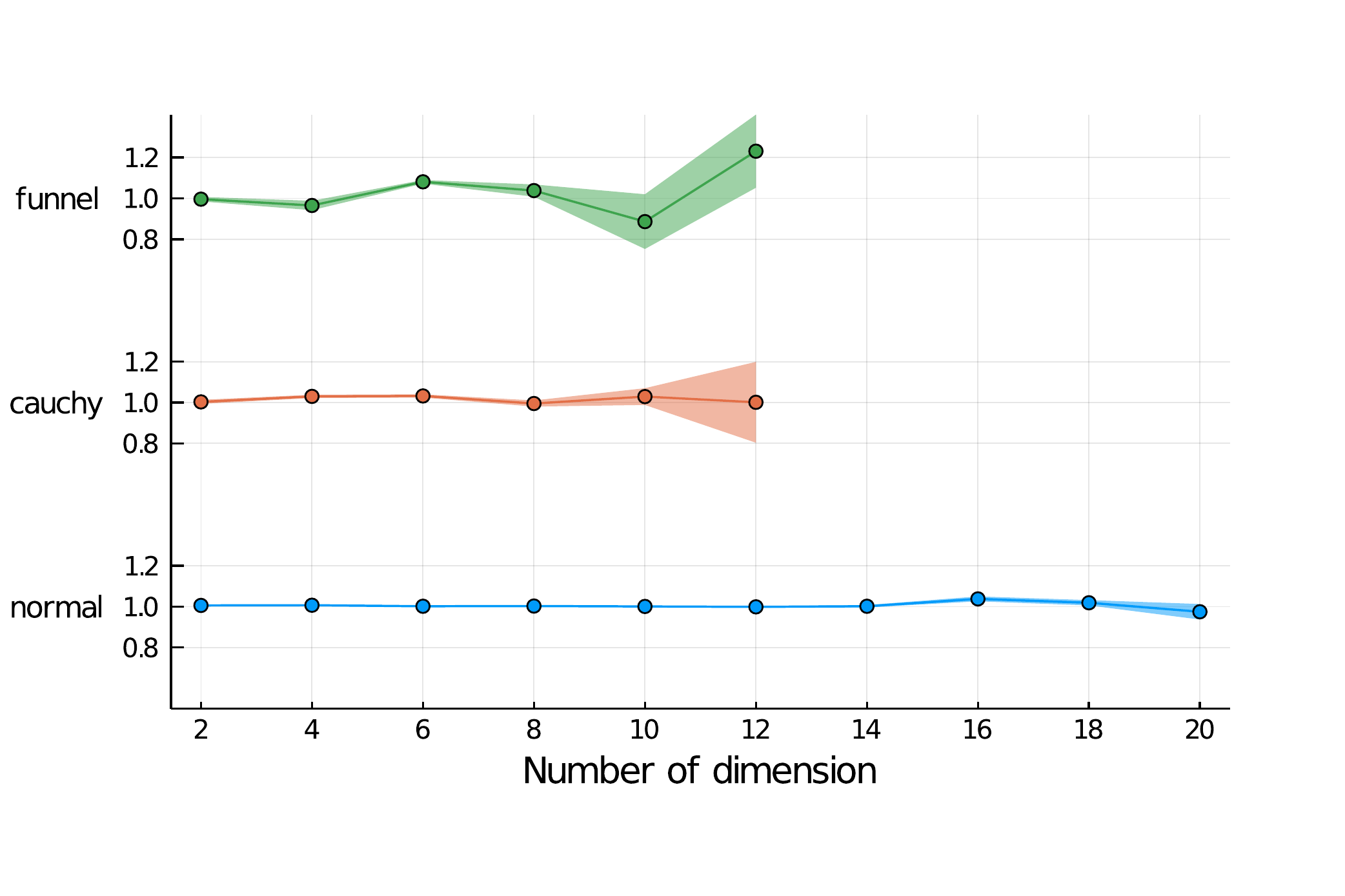}	
	\caption{The integral calculated using AHMI for the normal, multi-modal Cauchy and funnel distribution between 2 and 20 dimensions. The colored areas represent the uncertainty provided by AHMI.}
	\label{fig:ahmi}
\end{figure}

\begin{figure}
	\centering
	\includegraphics[width=0.7\textwidth]{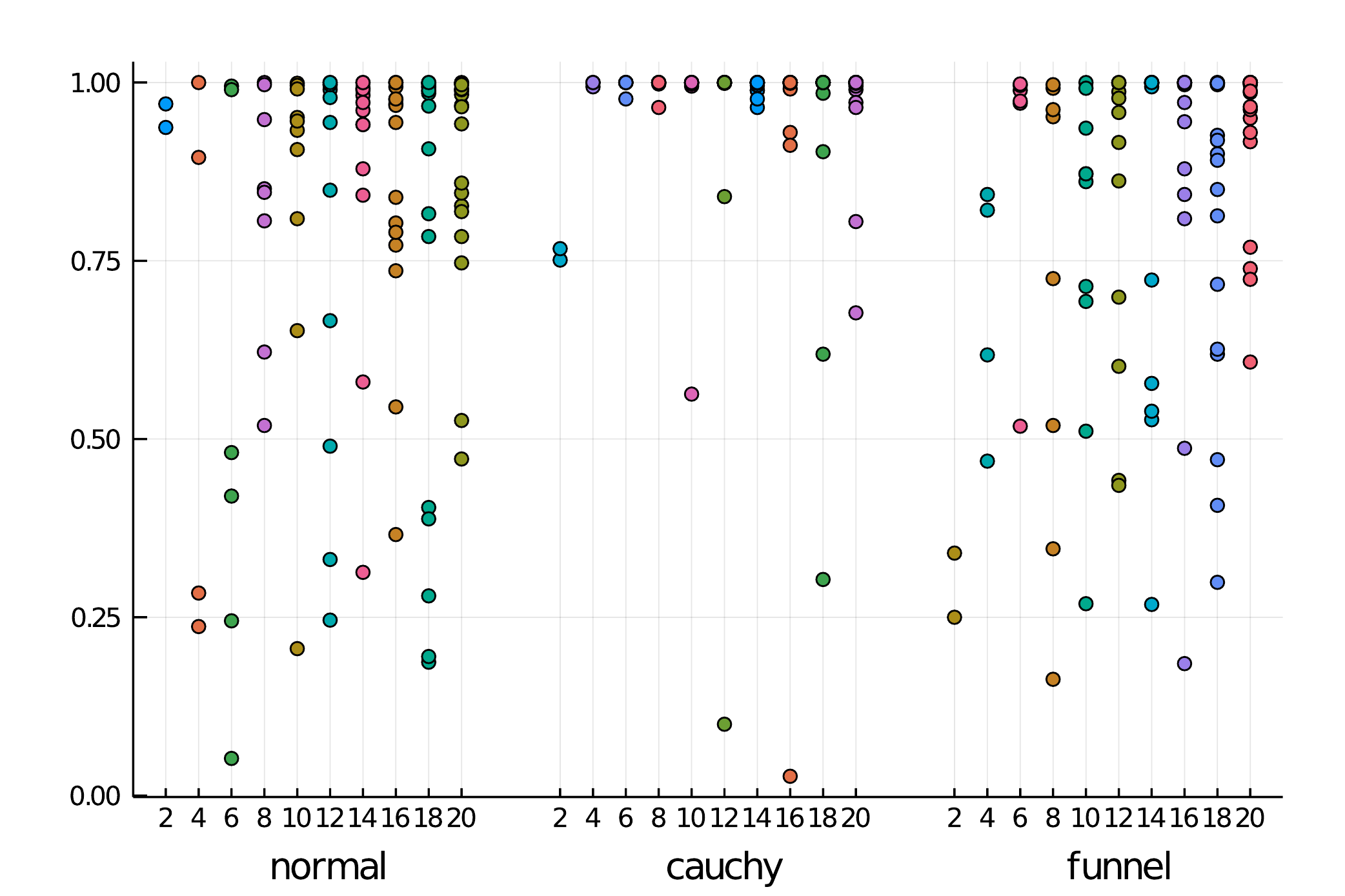}	
	\caption{The KS test p-values calculated for each marginal for the normal, multi-modal Cauchy and funnel distribution between 2 and 20 dimensions. The horizontal axis indicates the number of dimensions, while the p-value is given on the vertical axis.}
	\label{fig:ks_test_values}
\end{figure}

The AHMI integral and KS test p-values are calculated for the test functions from 2 up to 20 dimensions.
Fig.~\ref{fig:ahmi} shows the integral values and their uncertainties. The integral of the multi-modal Cauchy and funnel distribution are calculated for up to 12 dimensions using AHMI, whereas the integral for the normal distribution is calculated for up to 20 dimensions~\footnote{For a higher number of dimensions, the AHMI algorithm cannot determine appropriate integration subvolumes and reports its inability to perform the integral.}.
In all cases where the AHMI algorithm is able to report an integral value, the result is compatible within the quoted uncertainty with the expected value.
The distribution of the KS test p-values for the test functions from 2 to 20 dimensions is shown in Fig.~\ref{fig:ks_test_values}.
The distributions of the p-values for the normal and funnel distribution are compatible with the expectation. The p-values for the Cauchy distribution are, similar to the 2 dimensional performance measures, closer to one due to higher correlation of the samples.
We have also executed the test suite for the HMC sampling algorithm. Here, we present results for the funnel distribution in 20 up to 35 dimensions.  

\begin{figure}
	\centering
	\includegraphics[width=0.7\textwidth]{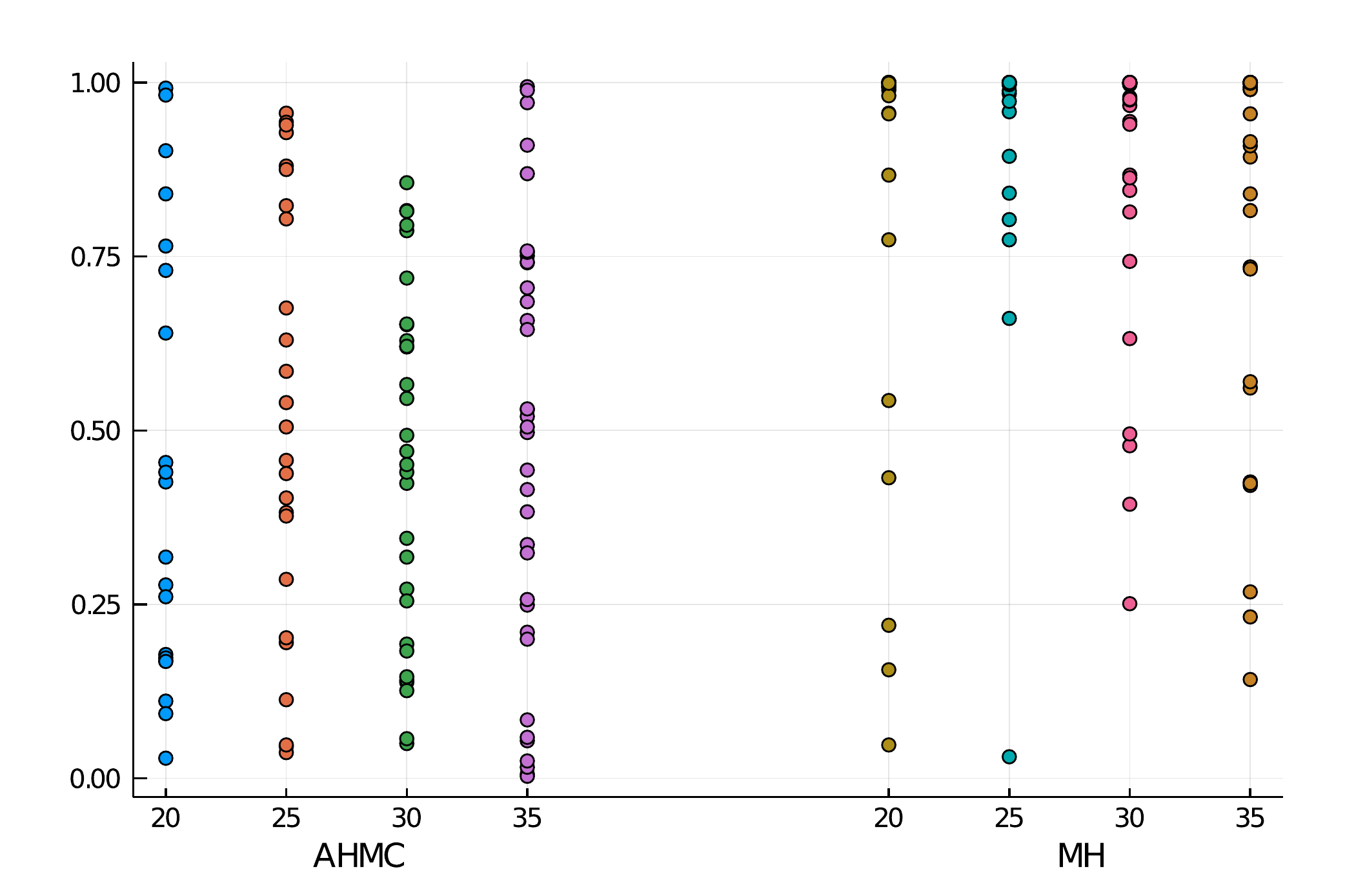}	
	\caption{KS p-values for all marginals of the sampled funnel distribution with 20 up to 35 dimensions, using both the HMC and the MH sampling algorithm. The horizontal axis indicates the number of dimensions, the p-value is given on the vertical axis.}
	\label{fig:ks_test_values_ahmc_mh}
\end{figure}

The KS p-values, shown in Fig.~\ref{fig:ks_test_values_ahmc_mh} follow an approximately flat distribution between 0 and 1, indicating that both sampling algorithms perform well. 

% -----------------------------------------------------------------------
\section{An extended example}
\label{sec:example}

In the following, we demonstrate the potential of \bat by solving a realistic problem of the type encountered in particle and astroparticle physics experiments; namely, fitting a model to a set of data and determining if a specific signal process is present in the data.

In the example, we imagine that we are searching for a rare phenomenon: e.g., a particular nuclear decay, which leaves a specific and well-defined signature in the experiment. The experiment itself comprises several different detectors that can measure the energy of an event and which are all sensitive to the signal in a limited energy window, for example from $0$ to $200$ \si{\keV}. The data we collect will come from two different sources, signal and background. We assume that while shielding measures are present that limit the detection of background events, we are not able to suppress them completely. 

In order to claim a discovery of the signal in this kind of experiment, it is not sufficient to detect events close to the energies predicted by the theory of the desired signal. Instead, the task is to make a statement on the probability of having detected signal events in the presence of background. This implies the comparison of two different models, namely the background-only (BKG) model, where we assume that no signal is present in the data and all events are due to background sources, and the signal-plus-background (S+BKG) model, where we assume that we detected events from both sources. We fit both of these models to the data and then compare them using a Bayes Factor.

\subsection{Data model}

As the experimental observable is set of energy values $\boldsymbol{E}$, we will formulate the model for signal and background processes in this quantity. We assume that the probability distribution for background events follows an exponential function characterized by a decay constant $\lambda$, i.e.

\begin{equation}
    p_{B}(E | \lambda) = \lambda \mathrm{e}^{-\lambda \cdot E} \, .
\end{equation}

The probability distribution for signal events follows a normal distribution with known mean value $\mu_S$ and also known standard deviation $\sigma_S$, i.e.

\begin{equation}
    p_{S}(E | \mu_S, \sigma_S) = \frac{1}{\sqrt{2 \pi} \cdot \sigma_S } \mathrm{e}^{-\frac{1}{2}\left( \frac{E - \mu_S}{\sigma_S} \right)^2}
\end{equation}

In the current example, we chose $\mu_S = 100$ \si{\keV} and $\sigma_S = 2.5$ \si{\keV}. Each detector will operate for a finite amount of time, $T_{i}$, also referred to as exposure. The total number of expected background events for detector $i$ is then

\begin{equation}
    \mu^B_{i} = T_{i} \cdot B_{i} \, ,
\end{equation}

where $B_{i}$ [$\textrm{counts} / \textrm{yr}$] estimates the background rate, i.e. the number of events per year of operation, assuming that the rate of background events does not change. 

Similarly, we can estimate the expected number of signal events in detector $i$ as

\begin{equation}
    \mu^S_{i} = T_{i} \cdot \epsilon_{i} \cdot S \, ,
\end{equation}

where $\epsilon_{i}$ is the efficiency of the detector to recognize the signal event and $S$ [$\textrm{events} / \textrm{yr}^{-1}$] is the signal rate and is representative of the signal strength.

Apart from modeling the data collected in the experiment, we might also need to model the detectors themselves. Suppose we use a total of five detectors in our experiment, but given that it takes time to build them, we start operating the detectors at different times resulting in different exposures. Since the detectors are produced one at a time and the manufacturer has time to refine the production process, the detection efficiency might be better for detector produced at a later stage. We can also assume that the background rates will not be exactly the same but they will be close to each other, since this quantity mostly depends on the properties of the detector material and the production process. In order to account for the correlation between the background rates of the different detectors, we assume that the individual background rates $B_{i}$ are randomly distributed according to a log-normal distribution, i.e.

\begin{equation}
    p(B_{i}) \sim \textrm{log-normal} \left(\mu_B, \sigma_B \right) \, .
\end{equation}

The log-normal distribution is a commonly used prior for non-negative parameters.

Since $p(B_{i})$ depends on $\mu_B$ and $ \sigma_B$, our prior will have a hierarchical, resp. layered structure. \bat allows the user to express hierarchical priors in straightforward fashion, the prior distribution of some model parameters can be expressed as a function of other model parameters.

Given the parameters $\mu_{B}$ and $\sigma_{B}$, the mean of the log-normal distribution is $m_{B} = e^{\mu_{B} + \frac{\sigma_{B}^2}{2}}$. In the following, we found it more intuitive to work with $m_{B}$ and then set $\mu_{B} = f(m_{B}, \sigma_{B})$. In our example, we assume five detectors with an exposure and efficiency given in Tab.~\ref{tab:exposure_table}.

\begin{table}
	\centering
	\caption{The exposure and efficiency of the fictional detectors.}
	\label{tab:exposure_table}
	\resizebox{0.3\textwidth}{!}{
	\begin{tabular}{|c | c | c |} \hline
 			 $i$		& Exposure $T_i$ [yr]	& Efficiency $\epsilon_i$ \\	\hline
           1 &          1.6 &      0.5 	\\
           2 &          1.3 &      0.6 	\\
           3 &          1.0 &      0.7 	\\
           4 &          0.7 &      0.8 	\\
           5 &          0.4 &      0.9 	\\ \hline
	\end{tabular}
	}
\end{table}

\subsection{Statistical model}

Since we have five detectors with different exposures and detection efficiencies, we split our data into five different datasets $\mathcal{D}_i$. The likelihood for the S+BKG model and a single dataset is then

\begin{equation}
    \mathcal{L}_i(\mathcal{D}_i | S, B_i) = \prod_{j = 1}^{N^{\textrm{obs}}_i} \frac{1}{\mu^B_i + \mu^S_i} \left[ \mu^B_i \lambda \mathrm{e}^{-\lambda E_j} + \mu^S_i \frac{1}{\sigma_S \sqrt{2 \pi}} \mathrm{e}^{-\frac{1}{2}\left( \frac{E_j - \mu_S}{\sigma_S} \right)^2} \right]
\end{equation}\\
where $N^{\textrm{obs}}_i$ is the number of events in dataset $\mathcal{D}_i$. The total likelihood is constructed as the product of all $\mathcal{L}_i$ weighted with the Poisson terms \cite{Beringer:2012bg}:

\begin{equation}
\label{likelihood_signal}
    \mathcal{L}(\left\{ \mathcal{D}_{i} \right\} | S, \left\{ B_{i} \right\}) = \prod_{i = 1}^5  \left[ \frac{\mathrm{e}^{-(\mu^B_i + \mu^S_i)}\left( \mu^B_i + \mu^S_i \right)^{N_i^{\textrm{obs}}}}{N_i^{\textrm{obs}} !} \cdot \mathcal{L}_i(\mathcal{D}_i | S, B_i) \right]
\end{equation}

We use the same likelihood for the BKG model, but with all $\mu^S_i$ set to zero. 

Apart from the likelihood, we also specify the priors for the free parameters of the model. These are the signal rate $S \sim \textrm{Uniform}(0, 10)$ $\textrm{yr}^{-1}$, the background rate parameters $m_B \sim \textrm{Uniform}(0, 5 \cdot 10^{-2})$ $\textrm{counts} / \textrm{yr}$ and $\sigma_B \sim \textrm{Uniform}(0.1, 1.0)$ $\textrm{counts} / \textrm{yr}$ as well as the decay constant $\lambda \sim \textrm{Uniform}(0, 100)$.

\subsection{Data and results}

\begin{figure}
    \begin{center}
    \includegraphics[width =0.65 \textwidth]{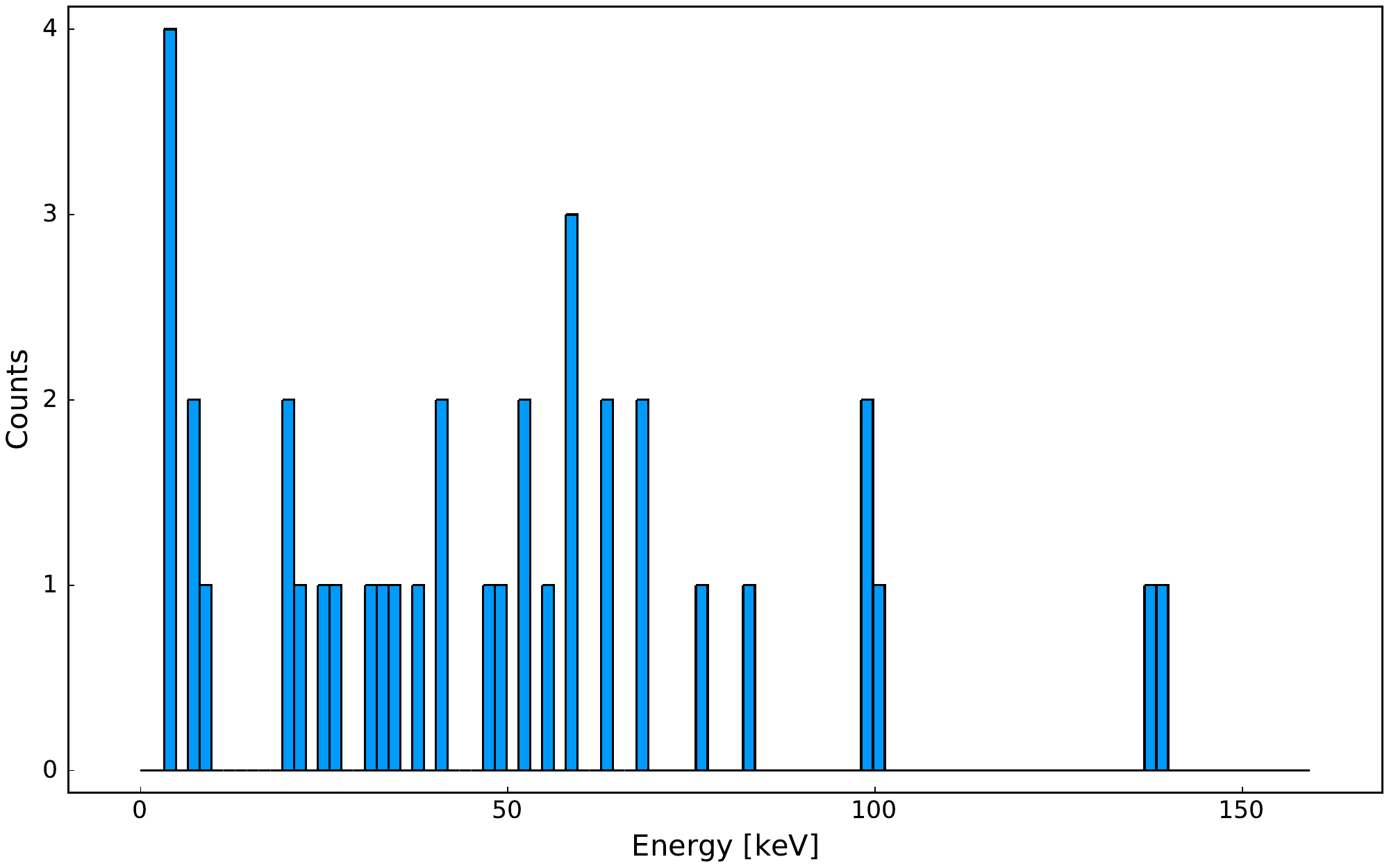}
    \end{center}
    \caption{Binned generated data}
\label{fig:physics_example_binned_data}
\end{figure}

The data for the analysis are generated synthetically. We choose a decay constant of $\lambda_{\textrm{true}} = 50$ with background rate parameters $m_B = 4.7$ and $\sigma_b = 0.5$. In addition, we include three signal events, i.e. $S=0.9375$. Figure~\ref{fig:physics_example_binned_data} shows the binned data. As can be seen, without knowing that there are three signal events at 100\,\si{\keV}, it would be very difficult to recognise them just by looking at the data.

\begin{figure}
    \begin{center}
    \includegraphics[width =0.8 \textwidth]{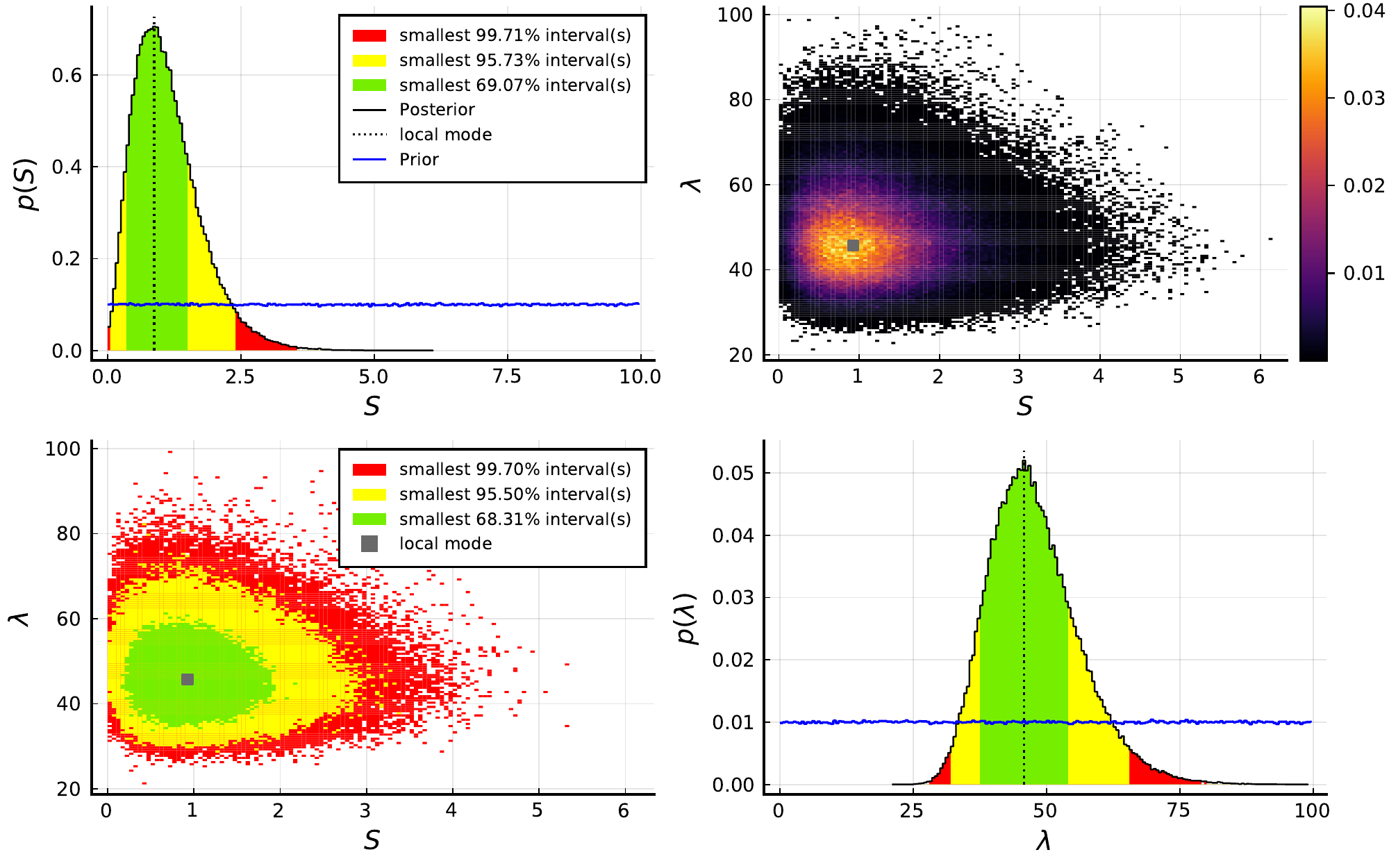}
    \end{center}
    \caption{Posterior distribution of the signal-rate $S$ and the background-rate $\lambda$. The blue line (in the upper-left and lower-right plot) shows the prior.}
\label{fig:prior_posterior}
\end{figure}

\begin{figure}
    \begin{center}
    \includegraphics[width =0.8 \textwidth]{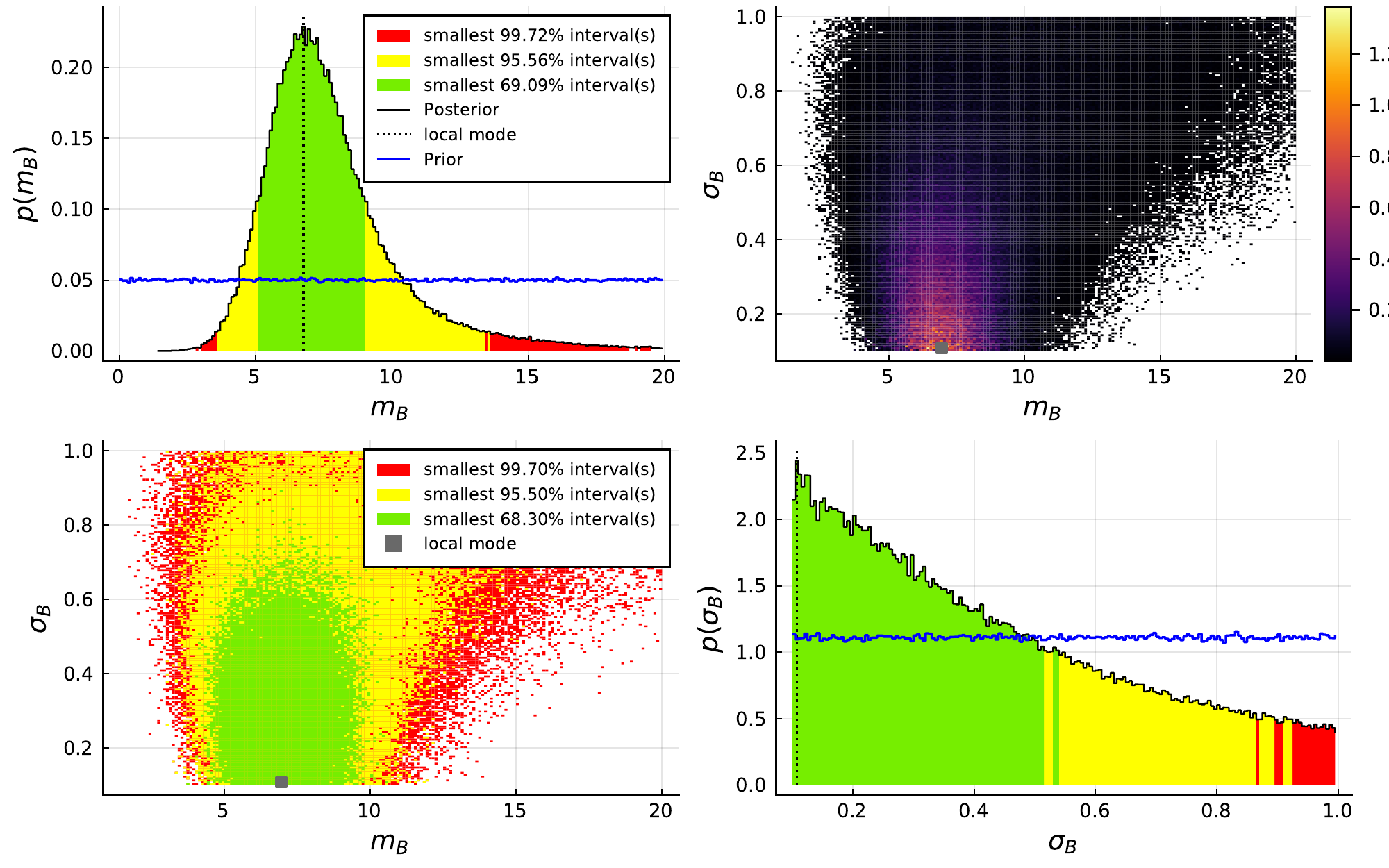}
    \end{center}
    \caption{Posterior distribution of the background rate mean $m_B$ and its sigma $\sigma_B$. The blue line (in the upper-left and lower-right plot) shows the prior.}
\label{fig:prior_posterior_hierarchical}
\end{figure}

In order to determine whether the model with or without signal should be preferred, we compute the evidences of the BKG and S+BKG models using AHMI and calculate the Bayes factor under the assumption of the same prior probability for the two models, i.e.

\begin{equation}
    \textrm{BF} = \frac{p(\textrm{S+BKG} | \mathcal{D})}{p(\textrm{BKG} | \mathcal{D})} = 3.4 \, ,
\end{equation}

which supports the claim that the data contains both signal and background events.

Having determined that our data does indeed contain signal, we look at the marginal posterior distributions in order to to check how well the fit reconstructs the parameters used to generate the synthetic data. With \bat the user can easily plot the results just like in Figure~\ref{fig:prior_posterior}, which shows both the 1D and 2D marginal posterior distributions for the signal rate $S$ and the background decay constant $\lambda$. In the marginalized distribution of a parameter, the mode is representative of the most likely scenario and inspecting the modes in Figure~\ref{fig:prior_posterior} we notice that $S$ peaks at 0.94 while $\lambda$ peaks at 47. Both modes are very close to the nominal values that were used in data generation.

Since we assume that there was a correlation between the background rates of the individual detectors, we examine the posterior of the model parameters that control the distribution of the $B_i$-s in Figure~\ref{fig:prior_posterior_hierarchical}. We notice that a mean background rate $m_B$ between 6 and 7 events per year is most likely. The spread of the posterior log-normal distribution will likely be small since the posterior of the parameter $\sigma_B$ exhibits an exponential decay peaking at 0.

\begin{figure}
    \begin{center}
    \includegraphics[width =0.8 \textwidth]{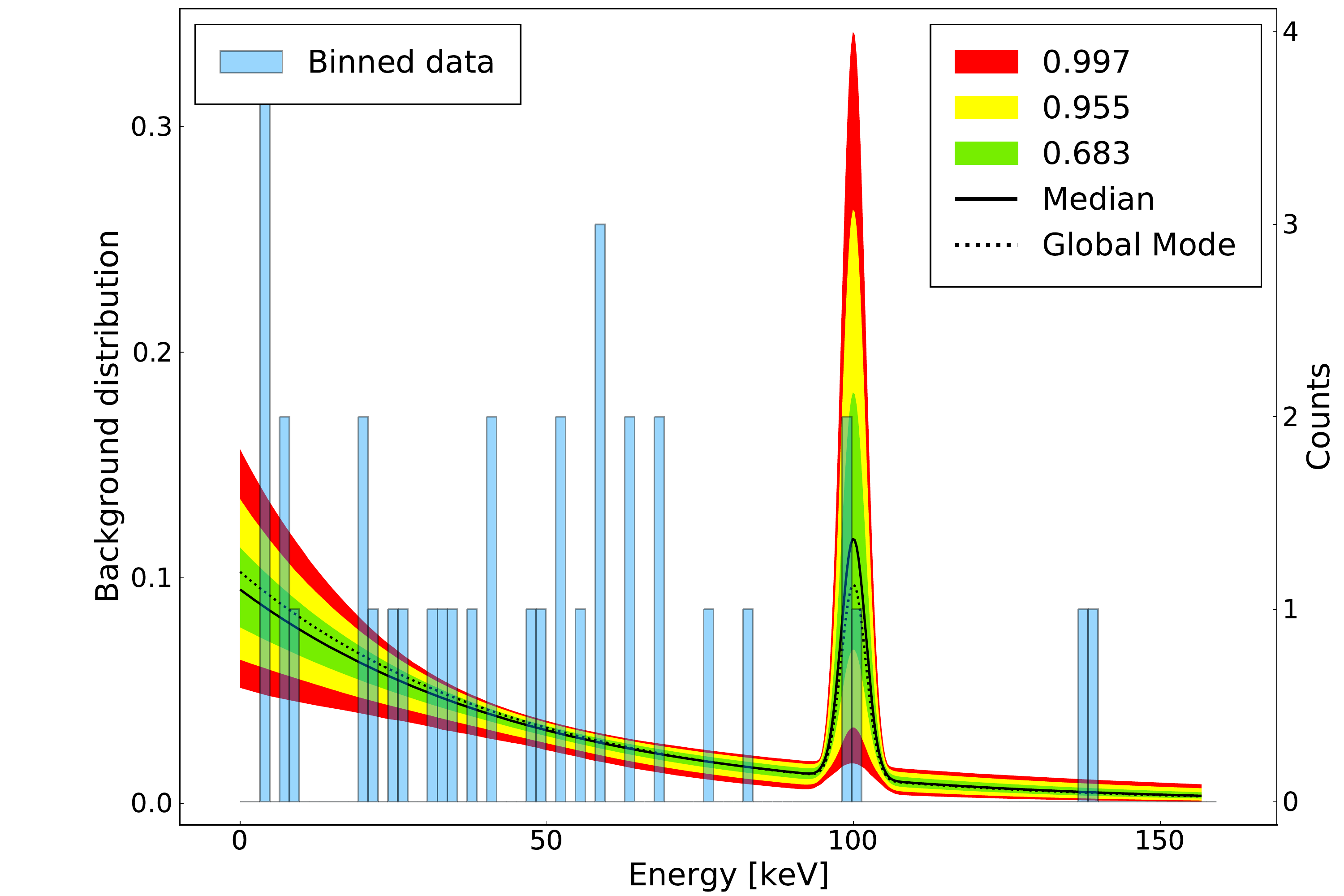}
    \end{center}
    \caption{Distribution of all data events, compared with the S+BKG model using best fit parameters and central quantiles according to the full posterior distribution.}
\label{fig:sum_detector_fit}
\end{figure}

Finally, we compare our S+BKG model with the data in Figure~\ref{fig:sum_detector_fit}.

% -----------------------------------------------------------------------
\section{Summary and outlook}
\label{sec:summary}

We have developed a platform-independent software package for Bayesian inference, \bat. \bat features a toolbox for numerical algorithms to perform calculations often encountered in Bayesian inference, in particular sampling, optimization and integration algorithms as well as flexible input/output routines. \bat also allows for interfacing with arbitrary custom codes, e.g. for the evaluation of complex models. We use the Julia programming language to provide a lightweight but powerful interface, parallel processing and automatic differentiation. We intend for the package to appeal to a wide user base, not constrained to a specific realm of science. The main application of \bat is to study models that are characterized by (numerically) complex likelihood functions. In this paper, we describe the design choices, the implemented algorithms, and the procedure to test the implementation. We also give a concrete physics example that demonstrates the capabilities of \bat. \bat has already seen first use in several scientific works~\cite{Bissmann:2019gfc,Bissmann:2019qcd,Stadnichuk:2005.02620,Caldwell:2005.11277,gerda:neutrino2020talk}.

For the future, we plan to extend the functionality available in \bat further, adding more algorithms, novel sampling schemes and multi-level parallelization.

% -----------------------------------------------------------------------
\section{Acknowledgements}

The authors would like to thank Tatyana Abramova for the fruitful discussions about Turchin's method of regularization and Hamiltonian Monte Carlo, and to thank Scott Hayashi for contributing to the \bat unit tests. C.G. is supported by the Studienstiftung des Deutschen Volkes. This project is supported by the Deutsche Forschungsgemeinschaft (DFG), project KR 4060/7-1, and by the European Union's Framework Programme for Research and Innovation Horizon 2020 (2014-2020) under the Marie Sklodowska-Curie Grant Agreement No.765710.

% -----------------------------------------------------------------------
\printbibliography

@article{Caldwell:2008fw,
      author         = "Allen Caldwell and Daniel Kollar and Kevin Kr{\"o}ninger",
      title          = "{BAT: The Bayesian Analysis Toolkit}",
      journal        = "Comput. Phys. Commun.",
      volume         = "180",
      year           = "2009",
      pages          = "2197-2209",
      doi            = "10.1016/j.cpc.2009.06.026",
      eprint         = "0808.2552",
      archivePrefix  = "arXiv",
      primaryClass   = "physics.data-an",
      SLACcitation   = "%%CITATION = ARXIV:0808.2552;%%"
}

@article{JSSv076i01,
   author = {Bob Carpenter and Andrew Gelman and Matthew Hoffman and Daniel Lee and Ben Goodrich and Michael Betancourt and Marcus Brubaker and Jiqiang Guo and Peter Li and Allen Riddell},
   title = {Stan: A Probabilistic Programming Language},
   journal = {Journal of Statistical Software, Articles},
   volume = {76},
   number = {1},
   year = {2017},
   issn = {1548-7660},
   pages = {1--32},
   doi = {10.18637/jss.v076.i01},
   url = {https://www.jstatsoft.org/v076/i01}
}

@Article{Lunn2000,
author="David J. Lunn
and Andrew Thomas
and Nicky Best
and David Spiegelhalter",
title="WinBUGS - A Bayesian modelling framework: Concepts, structure, and extensibility",
journal="Statistics and Computing",
year="2000",
volume="10",
number="4",
pages="325--337",
issn="1573-1375",
doi="10.1023/A:1008929526011",
url="https://doi.org/10.1023/A:1008929526011"
}

@Manual{R2017,
     title = {R: A Language and Environment for Statistical Computing},
     author = {{R Core Team}},
     organization = {R Foundation for Statistical Computing},
     address = {Vienna, Austria},
     year = {2017},
     url = {https://www.R-project.org/},
   }

@book{DAgostini2003,
      author = "Giulio D'Agostini",
      title = " Bayesian Reasoning in Data Analysis: A Critical Introduction",
      publisher = "World Scientific",
      year = " 2003"
}

@book{Hartigan1983,
      author = "John A. Hartigan",
      title = "Bayes Theory",
      publisher = "Springer New York",
      year = "1983"
}

@book{Jaynes2003,
      author = "Edwin T. Jaynes and G. Larry Bretthorst",
      title = "Probability Theory: The Logic of Science",
      publisher = "Cambridge University Press",
      year = "2003"
}

@book{Kendall1994,
      author = "Maurice George Kendall and others",
      title = "Kendall's Advanced Theory of Statistics: Bayesian Inference",
      publisher = "Hodder Arnold",
      year = "1994"
}

@book{MacKay2003,
      author = "David MacKay",
      title = "Information Theory, Inference and Learning Algorithms",
      publisher = "Cambridge University Press",
      year = "2003"
}

@book{Sivia2006,
      author = "Devinderjit Sivia and John Skilling",
      title = "Data analysis: a Bayesian tutorial",
      publisher = "Oxford University Press",
      year = "2006"
}

@article{Ullio:2016kvy,
      author         = "Piero Ullio and Mauro Valli",
      title          = "{A critical reassessment of particle Dark Matter limits
                        from dwarf satellites}",
      year           = "2016",
      eprint         = "1603.07721",
      archivePrefix  = "arXiv",
      primaryClass   = "astro-ph.GA",
      SLACcitation   = "%%CITATION = ARXIV:1603.07721;%%"
}

@article{Erdmann:2013rxa,
      author         = "Johannes Erdmann and others",
      title          = "{A likelihood-based reconstruction algorithm for
                        top-quark pairs and the KLFitter framework}",
      journal        = "Nucl. Instrum. Meth.",
      volume         = "A748",
      year           = "2014",
      pages          = "18",
      xdoi            = "10.1016/j.nima.2014.02.029",
      eprint         = "1312.5595",
      archivePrefix  = "arXiv",
      primaryClass   = "hep-ex",
      SLACcitation   = "%%CITATION = ARXIV:1312.5595;%%"
}

@article{Ghosh:2015wiz,
      author         = "Diptimoy Ghosh and Matteo Salvarezza and Fabrizio Senia",
      title          = "{Extending the Analysis of Electroweak Precision
                        Constraints in Composite Higgs Models}",
      year           = "2015",
      eprint         = "1511.08235",
      archivePrefix  = "arXiv",
      primaryClass   = "hep-ph",
      SLACcitation   = "%%CITATION = ARXIV:1511.08235;%%"
}

@inproceedings{Ciuchini:2014dea,
      author         = "Marco Ciuchini and others",
      title          = "{Update of the electroweak precision fit, interplay with
                        Higgs-boson signal strengths and model-independent
                        constraints on new physics}",
      booktitle      = "{International Conference on High Energy Physics 2014
                        (ICHEP 2014) Valencia, Spain, July 2-9, 2014}",
      xurl            = "https://inspirehep.net/record/1324391/files/arXiv:1410.6940.pdf",
      year           = "2014",
      eprint         = "1410.6940",
      archivePrefix  = "arXiv",
      primaryClass   = "hep-ph",
      SLACcitation   = "%%CITATION = ARXIV:1410.6940;%%"
}

@article{Ciuchini:2013pca,
      author         = "Marco Ciuchini and Enrico Franco and Satoshi Mishima
                        and Luca Silvestrini",
      title          = "{Electroweak Precision Observables, New Physics and the
                        Nature of a 126 GeV Higgs Boson}",
      journal        = "JHEP",
      volume         = "08",
      year           = "2013",
      pages          = "106",
      xdoi            = "10.1007/JHEP08(2013)106",
      eprint         = "1306.4644",
      archivePrefix  = "arXiv",
      primaryClass   = "hep-ph",
      SLACcitation   = "%%CITATION = ARXIV:1306.4644;%%"
}

@inproceedings{deBlas:2014ula,
      author         = "Jorge de~Blas and others",
      title          = "{Global Bayesian Analysis of the Higgs-boson Couplings}",
      booktitle      = "{International Conference on High Energy Physics 2014
                        (ICHEP 2014) Valencia, Spain, July 2-9, 2014}",
      xurl            = "https://inspirehep.net/record/1322562/files/arXiv:1410.4204.pdf",
      year           = "2014",
      eprint         = "1410.4204",
      archivePrefix  = "arXiv",
      primaryClass   = "hep-ph",
      SLACcitation   = "%%CITATION = ARXIV:1410.4204;%%"
}

@article{Luongo:2015zgq,
      author         = "Orlando Luongo and Pisani, Giovanni Battista and Troisi,
                        Antonio",
      title          = "{Cosmological degeneracy versus cosmography: a
                        cosmographic dark energy model}",
      year           = "2015",
      eprint         = "1512.07076",
      archivePrefix  = "arXiv",
      primaryClass   = "gr-qc",
      SLACcitation   = "%%CITATION = ARXIV:1512.07076;%%"
}

@article{Rappold:2015una,
      author         = "Christophe Rappold and others",
      title          = "{Hypernuclear production cross section in the reaction of
                        $^6Li$ + $^{12}C$ at 2A GeV}",
      journal        = "Phys. Lett.",
      volume         = "B747",
      year           = "2015",
      pages          = "129",
      xdoi            = "10.1016/j.physletb.2015.05.059",
      SLACcitation   = "%%CITATION = PHLTA,B747,129;%%"
}

@article{Kroeninger:2014bwa,
      author         = "Kr{\"o}ninger, Kevin and Schumann, Steffen and Willenberg,
                        Benjamin",
      title          = "{(MC)**3 -- a Multi-Channel Markov Chain Monte Carlo
                        algorithm for phase-space sampling}",
      journal        = "Comput. Phys. Commun.",
      volume         = "186",
      year           = "2015",
      pages          = "1",
      xdoi            = "10.1016/j.cpc.2014.08.024",
      eprint         = "1404.4328",
      archivePrefix  = "arXiv",
      primaryClass   = "hep-ph",
      SLACcitation   = "%%CITATION = ARXIV:1404.4328;%%"
}

@article{Caldwell:2014eca,
      author         = "Caldwell, Allen and Liu, Chang",
      title          = "{Target Density Normalization for Markov Chain Monte
                        Carlo Algorithms}",
      year           = "2014",
      eprint         = "1410.7149",
      archivePrefix  = "arXiv",
      primaryClass   = "physics.data-an",
      SLACcitation   = "%%CITATION = ARXIV:1410.7149;%%"
}

@article{Gelman-Rubin:1992gr,
      author         = "Gelman, Andrew and Rubin, Donal B.",
      title          = "{Inference from Iterative Simulation Using Multiple Sequences}",
      journal        = "Statistical Science",
      volume         = "07",
      number         = "04",
      year           = "1992",
      pages          = "503-511",
      xdoi            = "10.1214/ss/1177011148"
}

@article{Brooks-Gelman:1998bg,
      author         = "Gelman, Andrew and Rubin, Donal B.",
      title          = "{Inference from Iterative Simulation Using Multiple Sequences}",
      journal        = "Statistical Science",
      volume         = "07",
      number         = "04",
      year           = "1992",
      pages          = "503-511",
      xdoi            = "10.1214/ss/1177011148"
}

@article{Beringer:2012bg,
      author         = "Beringer, Juerg",
      title          = "{Review of Particle Physics}",
      journal        = "Physical Review D",
      volume         = "86",
      number         = "010001",
      year           = "2012",
      xdoi            = "10.1103/PhysRevD.86.010001"
}

@article{Bevan:2014tha,
      author         = "Bevan, A. J. and others",
      title          = "{The UTfit collaboration average of D meson mixing data:
                        Winter 2014}",
      collaboration  = "UTfit",
      journal        = "JHEP",
      volume         = "03",
      year           = "2014",
      pages          = "123",
      xdoi            = "10.1007/JHEP03(2014)123",
      eprint         = "1402.1664",
      archivePrefix  = "arXiv",
      primaryClass   = "hep-ph",
      SLACcitation   = "%%CITATION = ARXIV:1402.1664;%%"
}

@misc{bat.jl,
    author = "Oliver Schulz and others",
    title = "BAT.jl",
    url = "doi:10.5281/zenodo.2587213"
}

@misc{ahmi,
    author = {Allen Caldwell and Philipp Eller and Vasyl Hafych and Rafael C. Schick and Oliver Schulz and Marco Szalay},
    title = {Integration with an Adaptive Harmonic Mean Algorithm},
    year = {2018},
    journal        = " To appear in IJMP",
    eprint = {arXiv:1808.08051},
}

@article{Metropolis:1953am,
    author = "Nicholas Metropolis and Arianna W. Rosenbluth and Marshall N. Rosenbluth and Augusta H. Teller",
    title = "{Equation of state calculations by fast computing machines}",
    doi = "10.1063/1.1699114",
    journal = "J. Chem. Phys.",
    volume = "21",
    pages = "1087--1092",
    year = "1953"
}

@article{JMLR:v15:hoffman14a,
  author  = {Matthew D. Hoffman and Andrew Gelman},
  title   = {The No-U-Turn Sampler: Adaptively Setting Path Lengths in Hamiltonian Monte Carlo},
  journal = {Journal of Machine Learning Research},
  year    = {2014},
  volume  = {15},
  number  = {47},
  pages   = {1593-1623},
  url     = {http://jmlr.org/papers/v15/hoffman14a.html}
}

@inproceedings{ge2018t,
  author    = {Hong Ge and
               Kai Xu and
               Zoubin Ghahramani},
  title     = {Turing: a language for flexible probabilistic inference},
  booktitle = {International Conference on Artificial Intelligence and Statistics,
               {AISTATS} 2018, 9-11 April 2018, Playa Blanca, Lanzarote, Canary Islands,
               Spain},
  pages     = {1682--1690},
  year      = {2018},
  url       = {http://proceedings.mlr.press/v84/ge18b.html},
}

@article{DUANE1987216,
title = "Hybrid Monte Carlo",
journal = "Physics Letters B",
volume = "195",
number = "2",
pages = "216 - 222",
year = "1987",
issn = "0370-2693",
doi = "https://doi.org/10.1016/0370-2693(87)91197-X",
url = "http://www.sciencedirect.com/science/article/pii/037026938791197X",
author = "Simon Duane and A.D. Kennedy and Brian J. Pendleton and Duncan Roweth",
}

@inbook{doi:10.1201/b10905-7,
 author = "Radford M. Neal",
 title = "MCMC Using Hamiltonian Dynamics",
 chapter = {chapter5},
 booktitle = "Handbook of Markov Chain Monte Carlo",
 publisher = "CRC Press",
 year = 2011,
 doi = {10.1201/b10905-7},
 URL = {https://www.routledgehandbooks.com/doi/10.1201/b10905-7}
 }

@ARTICLE{2017arXiv170102434B,
       author = {{Betancourt}, Michael},
        title = "{A Conceptual Introduction to Hamiltonian Monte Carlo}",
      journal = {arXiv e-prints},
         year = 2017,
        month = jan,
          eid = {arXiv:1701.02434},
        pages = {arXiv:1701.02434},
archivePrefix = {arXiv},
       eprint = {1701.02434},
 primaryClass = {stat.ME},
       adsurl = {https://ui.adsabs.harvard.edu/abs/2017arXiv170102434B}
}

@article{PYMC,
author= "John Salvatier, Thomas Wiecki, Christopher Fonnesbeck",
title="Probabilistic programming in Python using PyMC3",
doi="10.7717/peerj-cs.55"
}

@article{DBLP:journals/corr/BezansonEKS14,
  author    = {Jeff Bezanson and
               Alan Edelman and
               Stefan Karpinski and
               Viral B. Shah},
  title     = {Julia: {A} Fresh Approach to Numerical Computing},
  journal   = {CoRR},
  volume    = {abs/1411.1607},
  year      = {2014},
  url       = {http://arxiv.org/abs/1411.1607},
  archivePrefix = {arXiv},
  eprint    = {1411.1607},
  bibsource = {dblp computer science bibliography, https://dblp.org}
}

@article{Zenger:52625,
      title = {Independently Extensible Solutions to the Expression  Problem},
      author = {Zenger, Matthias and Odersky, Martin},
      year = {2004},
      url = {http://infoscience.epfl.ch/record/52625},
}

@article{RevelsLubinPapamarkou2016,
    title = {Forward-Mode Automatic Differentiation in {J}ulia},
   author = {Jarrett Revels and Miles Lubin and Theodore Papamarkou},
  journal = {arXiv:1607.07892 [cs.MS]},
     year = {2016},
     url = {https://arxiv.org/abs/1607.07892}
}

@article{Innes:2018zygote,
  author    = {Michael Innes},
  title     = {Don't Unroll Adjoint: Differentiating SSA-Form Programs},
  journal   = {CoRR},
  volume    = {abs/1810.07951},
  year      = {2018},
  url       = {http://arxiv.org/abs/1810.07951},
  archivePrefix = {arXiv},
  eprint    = {1810.07951},
  bibsource = {dblp computer science bibliography, https://dblp.org}
}

@article{besard2018juliagpu,
  author        = {Besard, Tim and Foket, Christophe and De Sutter, Bjorn},
  title         = {Effective Extensible Programming: Unleashing {Julia} on {GPUs}},
  journal       = {IEEE Transactions on Parallel and Distributed Systems},
  year          = {2018},
  doi           = {10.1109/TPDS.2018.2872064},
  ISSN          = {1045-9219},
  archivePrefix = {arXiv},
  eprint        = {1712.03112},
  primaryClass  = {cs.PL},
}

@article{Mogensen2018,
  doi = {10.21105/joss.00615},
  url = {https://doi.org/10.21105/joss.00615},
  year = {2018},
  publisher = {The Open Journal},
  volume = {3},
  number = {24},
  pages = {615},
  author = {Patrick K. Mogensen and Asbj{\o}rn N. Riseth},
  title = {Optim: A mathematical optimization package for Julia},
  journal = {Journal of Open Source Software}
}

@article{Salmon2011,
  doi = {10.1145/2063384.2063405},
  url = {https://doi.org/10.1145/2063384.2063405},
  year = {2011},
  articleno = {16},
  pages = {1-12},
  author = {John K. Salmon and Mark A Moraes and Ron O. Dror and David E. Shaw},
  title = {Parallel random numbers: As easy as 1, 2, 3},
  journal = {SC '11}
}

@article{Hahn:2004cuba,
  author = {Thomas Hahn},
  title = {Cuba - a library for multidimensional numerical integration},
  journal = {Computer Physics Communications},
  volume = {168},
  number = {2},
  year = {2005},
  pages = {78-95}
}

@article{Geyer:1992,
  author = {Charles J. Geyer},
  title = {Practical Markov Chain Monte Carlo},
  journal = {Statistical Science},
  volume = {7},
  number = {4},
  year = {1992},
  pages = {473-483}
}

@article{MadrasSokal:1988,
  author = {Neal Madras and Alan D. Sokal},
  title = {The pivot algorithm: A highly efficient Monte Carlo method for the self-avoiding walk},
  journal = {J Stat. Phys.},
  volume = {50},
  number = {},
  year = {1988},
  pages = {109-186}
}

@article{Nelder1965ASM,
  title={A Simplex Method for Function Minimization},
  author={John A. Nelder and Ronald Mead},
  journal={Comput. J.},
  year={1965},
  volume={7},
  pages={308-313}
}

@article{LBFGS1998,
title = "On the limited memory BFGS method for large scale optimization",
author = "Liu, {Dong C.} and Jorge Nocedal",
year = "1989",
doi = "10.1007/BF01589116",
volume = "45",
pages = "503--528",
journal = "Mathematical Programming",
issn = "0025-5610",
publisher = "Springer-Verlag GmbH and Co. KG",
number = "1-3",
}

@misc{MITLicense,
  title = {The MIT License},
  howpublished = {\url{https://opensource.org/licenses/MIT}},
  note = {Accessed: 2020-07-23}
}

@article{Bissmann:2019gfc,
    author = "Bi{\ss}mann, Stefan and Erdmann, Johannes and Grunwald, Cornelius and Hiller, Gudrun and Kr{\"o}ninger, Kevin",
    title = "{Constraining top-quark couplings combining top-quark and $\boldsymbol{B}$ decay observables}",
    eprint = "1909.13632",
    archivePrefix = "arXiv",
    primaryClass = "hep-ph",
    reportNumber = "DO-TH 19/17",
    doi = "10.1140/epjc/s10052-020-7680-9",
    journal = "Eur. Phys. J. C",
    volume = "80",
    number = "2",
    pages = "136",
    year = "2020"
}

@article{Bissmann:2019qcd,
    author = "Bi{\ss}mann, Stefan and Erdmann, Johannes and Grunwald, Cornelius and Hiller, Gudrun and Kr{\"o}ninger, Kevin",
    title = "{Correlating uncertainties in global analyses within SMEFT matters}",
    eprint = "1912.06090",
    archivePrefix = "arXiv",
    primaryClass = "hep-ph",
    month = "12",
    year = "2019"
}

@misc{Stadnichuk:2005.02620,
Author = {Egor Stadnichuk and Tatyana Abramova and Mikhail Zelenyi and Alexander Izvestnyy and Alexander Nozik and Vladimir Palmin and Ivan Zimovets},
Title = {Prototype of a segmented scintillator detector for particle flux measurements on spacecraft},
Year = {2020},
Eprint = {arXiv:2005.02620},
}

@misc{Caldwell:2005.11277,
Author = {Allen Caldwell and Vasyl Hafych and Oliver Schulz and Lolian Shtembari},
Title = {Infections and Identified Cases of COVID-19 from Random Testing Data},
Year = {2020},
Eprint = {arXiv:2005.11277},
}

@misc{gerda:neutrino2020talk,
  author       = {Yoann Kermaidic and others},
  title        = {{GERDA, Majorana and LEGEND - towards a background-free ton-scale Ge76 experiment}},
  month        = jul,
  year         = 2020,
  publisher    = {Zenodo},
  doi          = {10.5281/zenodo.3959593}
}

@article{RGG97,
    author = "Gareth O. Roberts and Andrew Gelman and W. R. Gilks",
    title = "{Weak Convergence and Optimal Scaling of Random Walk Metropolis Algorithms}",
    journal = "Ann.Appl.Probab.",
    volume = "7.1",
    pages = "110-120",
    year = "1997"
}

@thesis{FrederikBeaujean,
    author = "Frederik Beaujean",
    title = "{A Bayesian analysis of rare \textit{B} decays with advanced Monte Carlo methods}",
    year = "2012"
}

@article{Caldwell:2017mqu,
      author         = "Caldwell, Allen and Merle, Alexander and Schulz, Oliver
                        and Totzauer, Maximilian",
      title          = "{Global Bayesian analysis of neutrino mass data}",
      journal        = "Phys. Rev.",
      volume         = "D96",
      year           = "2017",
      number         = "7",
      pages          = "073001",
      doi            = "10.1103/PhysRevD.96.073001",
      eprint         = "1705.01945",
      archivePrefix  = "arXiv",
      primaryClass   = "hep-ph",
      reportNumber   = "MPP-2017-79",
      SLACcitation   = "%%CITATION = ARXIV:1705.01945;%%"
}

@article{Agostini:2017jim,
      author         = "Agostini, Matteo and Benato, Giovanni and Detwiler,
                        Jason",
      title          = "{Discovery probability of next-generation neutrinoless double-\ensuremath{\beta} decay experiments}",
      journal        = "Phys. Rev.",
      volume         = "D96",
      year           = "2017",
      number         = "5",
      pages          = "053001",
      doi            = "10.1103/PhysRevD.96.053001",
      eprint         = "1705.02996",
      archivePrefix  = "arXiv",
      primaryClass   = "hep-ex",
      SLACcitation   = "%%CITATION = ARXIV:1705.02996;%%"
}

\end{document}